\documentclass[apj]{emulateapj}
\usepackage{amssymb,amsmath}
\usepackage{natbib}

\newcommand{\im}{\overline{m}}

\shorttitle{The OVRO 40-m Blazar Monitoring Program}

\shortauthors{Richards et al.}

\submitted{Submitted to \apjs.}

\begin{document}

\title{Blazars in the \emph{Fermi} Era: The OVRO 40-m Telescope Monitoring Program}

\author{Joseph~L. Richards$^*$}
\email[$^*$]{joey@caltech.edu}
\author{Walter Max-Moerbeck}
\author{Vasiliki Pavlidou$^\dag$}
\thanks{$^\dag$Einstein fellow.}
\author{Oliver~G. King}
\author{Timothy~J. Pearson}
\author{Anthony~C.~S. Readhead}
\author{Rodrigo Reeves}
\author{Martin~C. Shepherd}
\author{Matthew~A. Stevenson}
\author{Lawrence~C.~Weintraub}
\affil{Cahill Laboratory of Astronomy and Astrophysics, California Institute of Technology, 1200 E California Blvd, Pasadena CA 91125}

\author{Lars Fuhrmann}
\author{Emmanouil Angelakis}
\author{J.~Anton Zensus}
\affil{Max-Planck-Institut-f\"{u}r-Radioastronomie, Auf dem H\"{u}gel 69, 53121 Bonn, Germany}

\author{Stephen~E. Healey}
\author{Roger~W. Romani}
\author{Michael~S. Shaw}
\affil{Department of Physics, Stanford University, Stanford CA 94305}

\author{Keith Grainge}
\affil{Astrophysics Group, Cavendish Laboratory, University of Cambridge, J~J~Thomson Ave, Cambridge CB3~0HE, UK}
\affil{Kavli Institute for Cosmology Cambridge, Madingley Road, Cambridge CB3~0HA, UK}

\author{Mark Birkinshaw}
\author{Katy Lancaster}
\author{Diana~M. Worrall}
\affil{H.~H.~Wills Physics Laboratory, University of Bristol, Tyndall Ave, Bristol BS8~1TL, UK}

\author{Gregory~B. Taylor}
\affil{Department of Physics and Astronomy, University of New Mexico, Albuquerque NM 87131}

\author{Garret Cotter}
\affil{Department of Astrophysics, University of Oxford, Keble Road, Oxford OX1~3RH, UK}

\author{Ricardo Bustos}
\affil{Departamento de Astronom\'{i}a, Universidad de Concepci\'{o}n, Casilla 160-C, Concepci\'{o}n, Chile}
\affil{Departamento de Astronom\'{i}a, Universidad de Chile, Casilla 36-D, Santiago, Chile}

\begin{abstract}\vspace*{3ex}
  The Large Area Telescope (LAT) aboard the \emph{Fermi} Gamma-ray Space Telescope provides an
  unprecedented opportunity to study gamma-ray blazars.  To capitalize on this opportunity,
  beginning in late 2007, about a year before the start of LAT science operations, we began a
  large-scale, fast-cadence 15~GHz radio monitoring program with the 40-m telescope at the Owens
  Valley Radio Observatory (OVRO).  This program began with the 1158 northern ($\delta>-20\degr$)
  sources from the Candidate Gamma-ray Blazar Survey (CGRaBS) and now encompasses over 1500 sources,
  each observed twice per week with about 4~mJy (minimum) and 3\% (typical) uncertainty.  Here, we
  describe this monitoring program and our methods, and present radio light curves from the first
  two years (2008 and 2009).  As a first application, we combine these data with a novel measure of
  light curve variability amplitude, the intrinsic modulation index, through a likelihood analysis
  to examine the variability properties of subpopulations of our sample.  We demonstrate that, with
  high significance (7-$\sigma$), gamma-ray-loud blazars detected by the LAT during its first
  11~months of operation vary with about a factor of two greater amplitude than do the gamma-ray
  quiet blazars in our sample.  We also find a significant (3-$\sigma$) difference between
  variability amplitude in BL~Lacertae objects and flat-spectrum radio quasars (FSRQs), with the
  former exhibiting larger variability amplitudes.  Finally, low-redshift ($z<1$) FSRQs are found to
  vary more strongly than high-redshift FSRQs, with 3-$\sigma$ significance.  These findings
  represent an important step toward understanding why some blazars emit gamma-rays while others,
  with apparently similar properties, remain silent.

\end{abstract}

\section{Introduction}
The rotating super-massive black holes that power active galactic nuclei (AGN) somehow accomplish
the remarkable feat of channeling energy derived from their rotation and accretion disks into two
relativistic jets oppositely-directed along the spin axis.  In spite of intensive observational
efforts over the last four decades, the detailed mechanism of this process has remained elusive,
and, although several processes have been suggested, we are still largely ignorant of the
composition of the jets and the forces that collimate them.  The first detailed collimation
mechanism to be proposed was that of a ``de Laval'' nozzle \citep{BR74}, which is now known to be a
likely cause of re-collimation on kpc scales, but not of the initial collimation, which, as revealed
by Very Long Baseline Interferometry (VLBI), clearly occurs on sub-parsec scales. Other early
theories, which involve magneto-hydrodynamic winds \citep{BZ77} and/or magnetic fields threading the
inner accretion disk \citep{BP82}, remain the most promising approaches to a full understanding of
the phenomenon.

An observational difficulty is that, except in a few cases (e.g. M87), radio observations, which
provide the most detailed images of active galaxies, only probe the relativistic jets down to the
point at which the jets become optically thick at a point some light-weeks or light-months from the
site of the original collimation. Higher-frequency observations are needed to probe deeper into the
jets, although interstellar scintillation observations do in some cases reveal the presence of radio
emission features in some AGN that are $\sim$~5--50~micro-arcseconds in extent
\citep{kedziora-chudczer_apj_1997, dennett-thorpe_discovery_2000, jauncey_origin_2000,
  rickett_interstellar_2002, rickett_interstellar_2006, lovell_masiv_2008}, which can be very
persistent~\citep{macquart_emergence_2007}.  These mysterious, very high brightness temperature
features are by no means understood, and are certainly of great interest.  At optical wavelengths,
rapid swings in the polarization position angle have been used to tie together flux density
variations at TeV energies and variations at millimeter wavelengths~\citep{marscher_nature_2008}. At
very high energies of hundreds of GeV to TeV, very rapid variations down to timescales of minutes
have been observed by the HESS, MAGIC and VERITAS
instruments~\citep[e.g.][]{aharonian_exceptional_2007, aharonian_simultaneous_2009,
  acciari_radio_2009, acciari_discovery_2010}.  Full three-dimensional (non-axisymmetric)
magnetohydrodynamic relativistic simulations are now being carried out that enable detailed
interpretation of the observations over the whole electromagnetic
spectrum\citep[e.g.][]{mckinney_stability_2009, penna_simulations_2010}.

Relativistic beaming introduces complications in observational studies of relativistic jets.  The
continuum emission is strongly beamed along the jet axis, introducing strong observational selection
effects. Those objects having jets that are aligned at a small angle to the line of sight are
collectively known as ``blazars.'' Small variations in the angle between the jet axis and the line
of sight result in a large range of observed properties, such as apparent luminosity, variability,
and energy spectrum. Strong boosting of the continuum synchrotron emission from the jet also
frequently swamps optical line emission, making it difficult or even impossible to obtain a redshift
for the source. As a result, blazars are subdivided into two classes: flat-spectrum radio quasars
(FSRQs) and BL~Lacertae objects (BL~Lacs). The former class contains blazars dominated by strong
broad emission lines while the latter class contains those blazars with spectra dominated by their
continuum emission, and hence weak, if any, emission lines and very weak absorption lines, or no
lines at all. The large variations in the energy spectrum make it difficult to study many blazars
over the whole electromagnetic spectrum. As a result the study of large, carefully-selected samples
is necessary to determine the physical processes and conditions of the parent population.  As
relativistically boosted emission can be detected even from high-redshift sources, any intrinsic
scatter in jet properties and scatter due to relativistic beaming is additionally convolved with any
cosmological evolution of the black holes giving rise to the jets and their environment. It is
therefore not surprising that the study of the population properties of relativistic jets has, to
this day, been sparse at best.

The launch of the \emph{Fermi} Gamma-ray Space Telescope in June of 2008 provides an unprecedented
opportunity for the systematic study of blazar jets~\citep{Atwood09}. Its Large Area Telescope (LAT)
observes the sky at energies between 100~MeV and a few hundred~GeV. In this energy range
relativistic particles can be probed through their inverse Compton emission in the case of
electron/positron jets~\citep[e.g.][]{DSM92,SBR94,BL95}, or a combination of pionic emission from
primaries and inverse Compton emission from cascade-produced leptonic secondaries in the case of
hadronic jets~\citep[e.g.][]{Mannheim93}.

Blazars comprise the most numerous class of extragalactic GeV sources associated with lower-energy
counterparts: the first-year \emph{Fermi} point source catalog (1FGL) contains 1451 sources, of
which 596 have been associated with blazars in the first AGN catalog (1LAC)~\citep{abdo_fermi_2010,
  abdo_first_2010}. As the LAT completes one survey of the whole sky every 3 hours, it can provide
continuous monitoring of \emph{all} gamma-ray bright blazars, although with variable cadence that
depends on the integration time necessary to detect each object (which can range from a single
satellite pass to many months, depending on the average flux density of the object and its activity
state).

The exact location of the gamma-ray emission region and its proximity to the central black hole
remain subjects of debate.  Two possible models of the GeV emission region are that this emission
comes from a ``gamma-sphere'' close to the base of the jet \citep{BL95}, or that it comes from the
same shocked regions that are responsible for the radio emission seen in VLBI observations much
further out in the jet \citep{Jor01}. If the former model is correct then the gamma-ray observations
might well provide evidence of the initial collimation mechanism.

The testing of models of the location, structure, and radiative properties of the gamma-ray emission
region in blazars requires, in addition to the \emph{Fermi} observations, supporting broadband
observations of likely gamma-ray sources in various activity states. Such multiwavelength efforts
can occur in two modes:
\begin{enumerate}
\item regular monitoring of a preselected, statistically complete sample of likely
  gamma-ray-bright objects, independent of their gamma-ray activity state; and
\item intensive observations of archetypal objects or objects exhibiting unusual behavior
\end{enumerate}
The blazar monitoring program we discuss here is focused on the first mode.  In anticipation of the
unique opportunities offered by the \emph{Fermi}-LAT sky monitoring at gamma-ray energies, three
years ago we began the bi-weekly monitoring of a large sample including 1158 likely gamma-ray loud
blazars, preselected according to uniform criteria, with the Owens Valley Radio Observatory (OVRO)
40-m telescope at 15~GHz.  We also apply our observations in studies of the second mode through
\emph{Fermi}-LAT multiwavelength campaigns for flaring
sources~\citep[e.g.][]{abdo_fermi/large_2009,fermi_3c279_2010}) and through collaboration with the
F-GAMMA project, a complementary effort representing the second mode, focused on radio and sub-mm
spectral monitoring of about 60 prominent sources~\citep{angelakis_f-gamma_2010,
  fuhrmann_simultaneous_2007}.

The sample that we are studying with the OVRO 40-m telescope is statistically well-defined and large
enough to allow for statistical analyses and comparisons of sub-samples. In addition, as the 40-m
telescope is dedicated full time to this project, the cadence is high enough to allow sampling of
the radio light curves on timescales comparable with those typically achieved by \emph{Fermi}-LAT
for bright gamma-ray blazars and in this sense the 40-m and \emph{Fermi}-LAT are ideally matched.
The combination of sample size and cadence is unprecedented, making this by far the largest
monitoring survey of radio sources that has been undertaken to the date of writing.

Data from this program, in combination with \emph{Fermi} observations, will allow us systematically
to derive the radio and radio/gamma ray observational properties of the blazar population, including
\begin{itemize}
\item the radio variability properties of the blazar population, their dependence on redshift,
  spectral classification, luminosity, and gamma-ray activity;
\item any differences between the radio properties of gamma-ray loud blazars and blazars with
  similar radio luminosity which have not been detected by \emph{Fermi};
\item the properties (e.g. significance of correlation and the length and sign of any time delays)
  of cross-correlations between radio and gamma-ray flares of gamma-ray loud blazars; and
\item the combination of radio properties, if one exists, that can predict the apparent gamma-ray
  luminosity of a blazar (which, in turn, could be used to derive blazar gamma-ray luminosity
  functions from radio luminosity functions).
\end{itemize}
Such a systematic study of radio and radio/gamma population properties should allow us to address a
series of long-standing questions on the physical properties of blazar jets, including the location,
structure, and radiative properties of the gamma-ray emission region, and the collimation,
composition, particle acceleration, and emission mechanisms in blazar jets.

In this paper we describe in detail the 40-m telescope 15~GHz monitoring program, we present results
from the first two years of the program (2008 and 2009), and we derive the variability amplitude
properties of the blazar population at 15 GHz.  Studies of other blazar population radio and
radio/gamma properties will be discussed in upcoming
publications~\citep[e.g.][]{max-moerbeck_ovro_2010,fluxfluxmethodology,fluxfluxLAT,fgamma_paper_0}.

The remainder of this paper is organized as follows. In \S~2, we discuss the telescope and receiver
and our measurement procedures.  In \S~3 we discuss the method of operation.  In \S~4 we discuss our
sample of sources and observing strategy.  In \S~5 we discuss data editing and calibration. Our
results, including light curves, the derivation of variability amplitudes for the blazar population,
and population studies using this analysis are presented in \S~6. We summarize and discuss our
conclusions in \S~7.


\section{Telescope and Receiver}

\subsection{Optics}
\label{sec:optics}
The OVRO ``40-m'' telescope is actually a 130-foot-diameter f/0.4 parabolic reflector with
approximately 1.1~mm rms surface accuracy on an altitude-azimuth mount.  A cooled receiver with two
symmetric off-axis corrugated horn feeds is installed at the prime focus. The telescope and receiver
combination produces a pair of approximately Gaussian beams ($157\arcsec $ FWHM), separated in
azimuth by 12\farcm 95. We refer to these two beams, somewhat arbitrarily, as the ``antenna'' beam
and the ``reference'' beam, or \emph{ant} and \emph{ref}. The receiver selects left-hand circular
polarization, so linearly polarized sources of all orientations may be monitored in total intensity.
By observing compact sources of known flux density, we find the aperture efficiency, $\eta_A \sim
0.25$.

This relatively low aperture efficiency is due to deliberate underillumination of the dish by the
feed---for monitoring observations of a large sample of objects aiming at flux density measurements
repeatable to within a few percent we must consider the trade-off between aperture efficiency and
pointing accuracy.  Underillumination of the antenna increases the beamwidth and reduces
susceptibility to pointing errors relative to more fully illuminating the antenna, in addition to
reducing exposure to thermal noise from ground spillover. Experience has shown that we are operating
at close to the optimum illumination for the most efficient use of the telescope: increasing the
aperture efficiency gains little because the thermal noise is already acceptably low for observing
the objects in our monitoring sample.  The on-source duty cycle is about 20\%---a factor of two in
efficiency is lost due to Dicke switching, and the rest of the lost time is due to slewing and
calibration.


When the 40-m telescope moves in elevation, gravity deforms its surface, changing the antenna gain
and focus location.  The entire feed/receiver system can be moved along the optical axis to adjust
the focus.  The optimum focus position as a function of elevation is measured about once per year,
but has not been found to vary significantly except when the receiver has been removed and
reinstalled during maintenance.  Due to thermal effects, the optimum focus also varies slightly between
day and night operation and with the angle between the telescope structure and the Sun.

\subsection{Receiver}
\begin{figure}
  \includegraphics[width=\columnwidth]{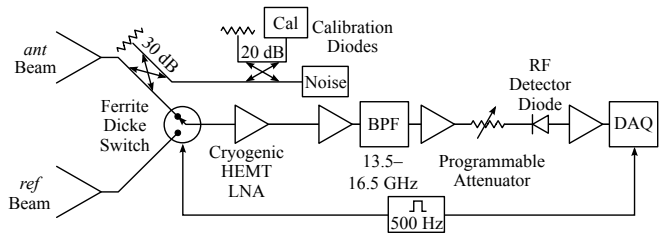}
  \caption[ovro40m_simple_blockdia]{Block diagram of the Ku-band
    receiver. \label{fig:receiver_block}}
\end{figure}

A block diagram of the receiver is shown in Figure~\ref{fig:receiver_block}.  The receiver
operates in the Ku band with a center frequency of 15.0~GHz, a 3.0~GHz bandwidth, and a
noise-equivalent reception bandwidth of 2.5~GHz.  The receiver noise temperature is about 30~K, and
the typical system noise temperature including CMB, atmospheric, and ground contributions is about
55~K.

In order to make the most efficient use of the telescope, a Dicke-switched dual-beam system is
used. A ferrite switch alternately selects between the \emph{ant} and \emph{ref\,} beams and delivers
the difference between the two, which is the switched power, i.e. the difference between the power
in \emph{ant} beam and the power in the \emph{ref} beam. We designate this power difference by $\xi$,
i.e.
\begin{equation}
  \xi = P_{\rm ant}-P_{\rm ref}
\end{equation}
Although Dicke-switching halves the time spent observing the object it is more efficient than using a
single-beam and scanning the telescope across the source
because the integrated signal from the source is higher and hence flux densities can be measured
faster, and in addition, as described below, Dicke-switching removes large
systematic errors.

The receiver front end consists of a cooled (T$\sim$80~K), low-loss ferrite RF Dicke switch followed
by a cryogenic (T$\sim$13~K) HEMT low-noise amplifier.  This is followed by additional
room-temperature amplifiers, a 13.5--16.5~GHz band definition filter, and an
electronically-controlled attenuator used to adjust the overall gain of the receiver.  The signal is
detected directly using a square law detector diode.  The detected signal is
digitized with a 16-bit analog-to-digital converter and then recorded.

%

From 2007 until 2008 November, several receiver calibrations were performed.  Beginning in 2008
November, approximately monthly calibrations were performed to monitor receiver performance.  These
calibrations included Y-factor measurements to characterize receiver temperature, sky dips to
measure atmospheric optical depth and to determine the ground spillover, calibration diode effective
temperature measurements, and observations of calibration sources to measure the aperture
efficiency.


\subsubsection{Measurement procedures}
\label{sec:measurement_procedures}
In typical radiometry observations on the 40-m telescope we use three procedures: (i) flux density
measurements, (ii) measurements of a calibration noise source, and (iii) pointing measurements on a
nearby bright source.  The receiver output voltage is integrated and digitally sampled at 1~kHz,
synchronously with the 500~Hz Dicke switching rate. Alternate millisecond samples are subtracted to
demodulate the Dicke switching, and the results are accumulated into 1-second averages.  In addition
to the demodulated outputs, the sum of the alternate samples, i.e. the total powers in both the
\emph{ant} and \emph{ref} beams are also recorded.

\subsubsection{Calibration diode}
\label{sec:cal_proc}
A pair of calibrated noise diodes, referred to as the ``noise'' and ``calibration'' diodes, are
connected to the main beam input via directional couplers to the Dicke switch.  These noise diodes
provide an excess noise ratio of ($31\pm 1$)~dB from 12--18~GHz with stability of about 0.001~dB/K.
The diodes provide two calibration levels---one of power comparable to the system temperature and
one attenuated to provide power comparable to the astronomical sources we are observing.  The
equivalent noise temperatures of the \emph{noise} and \emph{calibration} diodes at the receiver input
are about 67~K and 1~K, respectively, stable to $\ll$ 1\%.



\subsection{Pointing} \label{sec:pointing}
The 40-m telescope is equipped with encoders on the azimuth and elevation shafts and with two orthogonal tilt meters
located in the teepee of the telescope in the alidade above the azimuth bearing.  These four sets of readings are combined
in a pointing model that generates encoder azimuth and zenith angle
offsets based on the requested position on the sky. The pointing model has 9
terms for the azimuth angle correction and 5 terms for the zenith angle correction,
\begin{eqnarray}
  \nonumber \Delta \phi_{model} & = & A_{1}\sin\theta + A_{2} + A_{3}\sin\phi\cos\theta \\
  &+& A_{4}\cos\phi\cos\theta + A_{5}\cos\theta + A_{6}\sin\phi\sin\theta \\
  \nonumber & + & A_{7}\cos\phi\sin\theta + A_{8}\sin\left(4\theta\right) + A_{9} T_{LR}\cos\theta
\end{eqnarray}
\begin{equation}
  \Delta\theta_{model}  =  Z_{1} + Z_{2}\sin\theta - Z_{3}\cos\phi + Z_{4}\sin\phi + Z_{5} T_{AF}
\end{equation}
Here, $\phi$ and $\theta$ are the requested azimuth and zenith angles,
$\Delta\phi_{model}$ and $\Delta\theta_{model}$ are the pointing model
corrections for the azimuth and zenith angles, $A_{i}$ and $Z_{i}$ are the
pointing model coefficients, and $T_{AF}$ and $T_{LR}$ are the aft-forward and
left-right tilt meter readings.

\begin{figure}
  \includegraphics[width=\columnwidth]{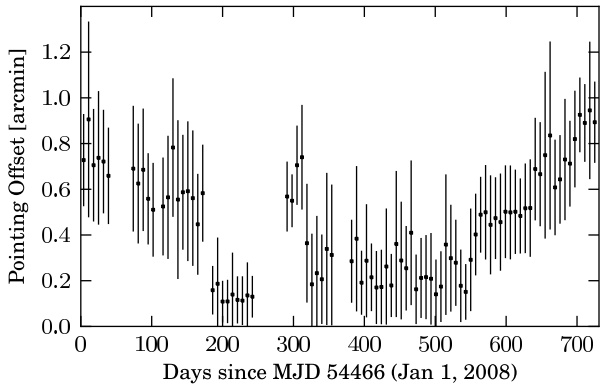}
  \caption[pointing_offset_weekly_means]{Residual error between the pointing model and the
    actual requested position, plotted in week-long bins.  The plotted data and errors are the
    weekly means and standard deviations of the pointing offsets measured by the pointing
    calibrations.}
  \label{figure:pointing_vs_t}
\end{figure}

We have found that the pointing model terms drift slowly with time.
Figure~\ref{figure:pointing_vs_t} shows the residual offset between the pointing model and the
actual requested position for 2008 and 2009.  The sharp steps in the average offset correspond to
adjustments in the pointing model.  We adjust the pointing model two to three times per year to
minimize the scatter in the offset and maintain an average offset less than about $0\farcm 5$ to
ensure accurate pointing.  Early in 2008 and at the end of 2009, the offset approached
$1\arcmin$, but because the scatter did not increase, there was no substantial impact on data
quality.

In addition to the pointing model correction, at least once per hour we measure the pointing offset
between a bright pointing calibrator and the model prediction.  This measures the effect of wind and
thermal loading. In early 2009, we determined that these pointing offsets have the accuracy we
require only at separations up to about 30\degr\ from the position where the pointing offset was
measured.  Because of this effect, after MJD~54906 (2009 March~16), care was taken when scheduling
to ensure that flux density measurements were always made at separations of less than 15\degr\ from
the pointing offset measurement.  Prior to this, no such limit was in place.  We have discarded flux
densities measured with a separation of more than 30\degr.

The pointing offset is measured by performing 3-point cross-scans of the calibrator in both azimuth
and zenith angle and fitting a fixed-width Gaussian beam profile to each axis to determine the
position of the peak.  A pointing offset measurement is considered invalid if its signal-to-noise
ratio is less than 2, or if the offset indicates that the peak was outside the span of the
cross-scan, $\pm$ FWHM/2.  Several iterations are attempted, moving the cross-scan center by up to
FWHM/2 after each attempt, allowing offsets less than the FWHM (157\arcsec) to be measured reliably.

\section{Method of Observing}
\label{sec:method}
\subsection{Double Switching}
\label{sec:double_switching}

The observing method we used follows closely that which we developed and described in detail in
\citet{readhead1989}, and also discussed in \citet{angelakis_multi-frequency_2009}.  To remove the
large varying total power signal and minimize the effects of the atmospheric fluctuations, ground
spillover, and gain fluctuations we use a ``double switching'' approach, with a Dicke switch
operating at 500~Hz, and azimuth switching in which we alternate the beams on the object of
interest.  The advantage of double switching is that large variable signals are eliminated and the
disadvantage is the loss of a factor of two in sensitivity---a factor of $\sqrt{2}$ lost through
observing the object only half the time, and another factor of $\sqrt{2}$ lost through the noise
introduced by subtracting off the reference field.

\subsubsection{Dicke Switching}

The most important benefit of Dicke switching is the removal of the large, slowly varying total
power signal, which is made up of contributions from ground, atmosphere, and receiver thermal noise.
Variations in the gain of the low noise amplifier cause variations in the large total power signal,
and in addition the signals themselves vary slowly with time and with the position of the
telescope. The resulting large variations in power limit the sensitivity of the receiving
system. Ground spillover, like gain variations, contributes directly to the system noise, but the
effect is difficult to quantify due to the complexity of the far sidelobes of the telescope
beam. Dicke switching removes these large slowly-varying signals.

A second benefit of Dicke switching is the reduction of noise due to the rapidly varying atmosphere
above the telescope. With a beam separation of 12\farcm 95, and for a water vapor scale height of
1.5~km, 75\% of the total mass of water vapor seen by the telescope lies in the overlapping portions
of the two beams.  This fraction does not change substantially with scale height, dropping only to
72\% (70\%) for a water vapor scale height of 2~km (2.5~km).  So Dicke switching reduces the effects
of the varying atmosphere by about a factor of 4.

A third benefit of Dicke switching is the relatively short observing time compared to the time
required to scan across a radio source with a single beam.  A detailed discussion of these benefits
is given in \citet{readhead1989}.

\subsubsection{Beam Switching and Flux Density Measurements}
\label{sec:flux_proc}

While Dicke switching does much to reduce the large error terms due to the atmosphere, the ground,
and gain fluctuations in the receiver, it does not remove linear drifts in any of these quantities
and the situation can be further improved by beam switching.  Beam switching in azimuth is optimum
because by maintaining a constant elevation we minimize changes to the atmospheric and ground
spillover signals and thereby maximize their cancellation.  We therefore adopt the same ``double
switching'' technique used by \citet{readhead1989}, in which we alternate the two beams on the
object of interest, and hence remove both the constant term and any linear drifts in the power from
these unwanted components of the signal.

The procedure we use for measuring flux densities is identical to that described in detail in
\citet{readhead1989} so we do not repeat all the details here, but give only a summary.  To begin
with the \emph{ref} beam is positioned on the source for 8 seconds, and the power difference,
$\xi_A$, is recorded. Then the \emph{ant} beam is positioned on the source for 8 seconds and the
power difference, $\xi_B$, is recorded.  With the \emph{ant} beam still on the source a second 8
second observation is then made and the difference, $\xi_C$ is recorded. Finally the \emph{ref} beam
is again positioned on the source for a final 8 second period and the difference, $\xi_D$ is
recorded.  Thus we spend a total of 32 seconds actually integrating on the source for each flux
density measurement.  Of course, slewing and settling times have to be allowed for at the beginning
of the $A$, $B$ and $D$ integrations, so that the total time required for the flux density
measurement is about one minute.

The corresponding flux density is given by
\begin{equation}\label{eq:flux}
  S_{15}= {\kappa \over 4} (\xi_{B}+\xi_{C}-\xi_{A}-\xi_{D})
\end{equation}
where $\kappa$ is the calibration factor required to turn digitizer units into Jy, and the
rms error is given by
\begin{equation}\label{eq:flux_err}
  \sigma_{15}= {\kappa \over 4}\sqrt{\sigma_{A}^{2} + \sigma_{B}^{2} + \sigma_{C}^{2} + \sigma_{D}^{2}}
\end{equation}
The calibration factor consists of a relative calibration factor that is computed for each measurement
(\S~\ref{sec:relative_cal}) and an absolute calibration factor (\S~\ref{sec:absolute_cal}).

The four measurements also contain interesting information on the stability of the instrument and,
more importantly, the atmosphere, during the observations. For each flux density measurement, we
therefore also compute two other quantities---one that we call the ``switched power,'' $\psi$,
given by
\begin{equation}\label{eq:swp}
  \psi={\kappa \over 4}\left(\xi_B+\xi_D-\xi_A-\xi_C\right)
\end{equation}
and the other that we call the ``switched difference,'' $\mu$, given by
\begin{equation}\label{eq:swd}
  \mu={\kappa \over 4}\left(\xi_C+\xi_D-\xi_A-\xi_B\right).
\end{equation}
Both $\psi$ and $\mu$ should be zero in the absence of gain or atmospheric drifts so we use these as
a way of estimating such variations in our error model (\S~\ref{sec:main_unc}) and to reject badly-contaminated measurements
(\S~\ref{sec:swd}).  The uncertainties in $\psi$ and $\mu$ are clearly given by
equation~(\ref{eq:flux_err}).

\subsection{Confusion}

For sources at galactic latitude $|b| > 10\degr $ most of the reference fields are empty, but there
are some objects that are contaminated by confusion introduced by other radio sources in the field.
Fortunately since we are observing bright sources confusion is not a problem. At 15.2~GHz,
\citet{waldram2010} report a differential source count $n(S)\approx 51
(S/\mathrm{Jy})^{-2.15}~\mathrm{Jy}^{-1}~\mathrm{sr}^{-1}$ with no deviation to a completeness limit
of 5.5~mJy.  Assuming that the effect of source clustering is negligible, the expected number of
confusing sources detected at or above a flux density limit $S_{c}$ in either the \emph{ant} or
\emph{ref} beam is
\begin{equation}\label{eq:num_confusing}
  N(S_{c}) = \int_{S_{c}}^{\infty}n(S)\Omega(S)\,dS
\end{equation}
where $\Omega(S)$ is the beam solid angle with antenna gain sufficient to detect a source of flux
density $S$ at the $S_{c}$ level.  For a beam-switched flux density measurement, the expected number
of confusing sources in the main or either reference beam is then $\nu=N(S_{c}) +
2N\left(S_{c}\sqrt{2}\right)$ where the confusion limit is higher in the reference beams results
because each is integrated only half as long as the main beam.  Considering the confusing sources to
be independently distributed among the observed fields via a Poisson process, the probability that a
beam-switched flux density measurement includes one or more confusing sources is $1-e^{-\nu}$.

Table~\ref{table:confusion} shows the probability of a confusing radio source lying in the main
field or either reference field of a single flux density measurement, as well as the expected number
of contaminated sources in our 1158 object sample.  Here, we have treated the \emph{ant} and
\emph{ref} beams as identical 157\arcsec\ FWHM Gaussian beams and neglected reference field
rotations with parallactic angle.  The latter approximation is justified because we observe sources
at approximately the same local sidereal time each cycle, limiting the parallactic angles at which
sources are observed.  Because only about 1.2\% of our sources are likely to be contaminated even at
10~mJy level ($\sim$~3\% of the median flux density of sources in our sample), we may safely ignore the
effects of confusion in our statistical analyses.

%
\begin{deluxetable}{c c c}
  \tablecaption{Confusion \label{table:confusion}}
  \tablehead{\colhead{Flux Density Limit (mJy)} & \colhead{Probability} & \colhead{Sources Affected\tablenotemark{a}}}
  \tablewidth{0pt}
  \startdata
  100 & $8.4 \times 10^{-4}$ & 1\\
  50  & $1.9 \times 10^{-3}$ & 2\\
  20  & $5.3 \times 10^{-3}$ & 6\\
  10  & $1.2 \times 10^{-2}$ & 14\\
  \enddata
  \tablenotetext{a}{Expected number of contaminated program sources, considering a source
    contaminated if a confusing source is found in the source field or either of its two reference
    fields.}
\end{deluxetable}

\section{Observations}

\subsection{Source Selection}

The selection of the core sample for our monitoring program was driven by three considerations.
First, since we are interested in the detailed study of the radio variability properties of the
blazar population and the dependence of these properties on other observables such as redshift, the
sample should be sufficiently large to allow division in subsamples (e.g. in redshift or luminosity
bins) with enough members to derive confidently the statistical properties in each.

Second, to allow for the evaluation of the confidence level of any correlations or variable
dependencies identified in our data through Monte-Carlo simulations, and the generalization of our
findings to the blazar population, the sample should be well-defined statistically, using uniform
and easily reproducible criteria.

Finally, one of the major goals of our monitoring program is the cross-correlation of 15 GHz light
curves with \emph{Fermi} gamma-ray light curves. For this reason we would like our sample to include
a large number of gamma-ray loud--blazars. On the other hand, we would also like to be able to
address the question of why some blazars are gamma-ray loud while other blazars, with apparently
similar properties, are not. For this reason we would like our sample to be
\emph{preselected}---before \emph{Fermi} data bias our understanding of what constitutes a likely
gamma-ray-loud blazar---and, ideally, to include a comparable number of blazars which are not
gamma-ray loud.

Blazars in the Candidate Gamma-Ray Blazar Survey (CGRaBS) satisfy all of the requirements
above~\citep{healey_cgrabs:all-sky_2008}. CGRaBS blazars were selected from a flat-spectrum parent
sample (complete to 65~mJy flux density at 4.8~GHz and radio spectral index $\alpha>-0.5$ where
$S\propto \nu^{\alpha}$) by a well-defined figure-of-merit criterion based on radio spectral index,
8.4~GHz radio flux density, and X-ray flux based on counts/s in the ROSAT All Sky Survey, to
resemble blazars that were detected by the Energetic Gamma-Ray Experiment Telescope (EGRET, the
precursor of \emph{Fermi}-LAT). The CGRaBS sample is a total of 1625 active galactic nuclei (AGN)
over the whole sky outside a $\pm 10\degr$ band around the galactic plane. This sample was compiled
before the launch of the \emph{Fermi} and was expected to contain a large fraction of the
extragalactic sources that would be detected by \emph{Fermi}-LAT.

The core sample for our monitoring program consists of the 1158 CGRaBS sources north of declination
$-20\degr$.  As published, our subset of the CGRaBS sample contains 812 FSRQs, 111 BL~Lacs, 53 radio
galaxies, and 182 objects without spectroscopic identification.  In our analysis we use redshifts
from the CGRaBS publication, which covered 81\% of the sample (100\% of FSRQs, 49\% of BL~Lacs).
Recent spectroscopy has improved the completeness of the sample to 886 FSRQs, 122 BL~Lacs, 60 radio
galaxies, and 88 objects without spectroscopic identification, with redshifts now available for
87.5\% of the sample (100\% of FSRQs; 53\% of BL~Lacs).  The median 15~GHz flux density for sources
in our sample ranged from about 20~mJy to 30~Jy with a median of 325~mJy during the observation
period described in this paper.

In 1LAC, 709 AGN were associated with 1FGL sources~\citep{abdo_first_2010}. Although continued
improvements in evaluating the probability of radio counterpart identification have caused some
source associations to vary in estimated significance and continued optical observations have
improved the completeness of the typing and redshifts, we adopt here the identifications published
in the 1LAC~\citep{abdo_first_2010}.  These identifications include 291 CGRaBS sources (221 of our
subset) that were associated with sources detected in gamma-rays by
\emph{Fermi}~\citep{abdo_first_2010}.  Of these, 263 (199 of our subset) were considered ``clean''
associations, meaning only one source was associated with the gamma-ray source and the association
probability was greater than 80\%. CGRaBS sources made up 44\% of the clean associations in the
first-year \emph{Fermi} AGN catalog. This number is thus far smaller than anticipated; in the
11-month 1LAC sample only $\sim$~16\% of the CGRaBS sources were detected and a large number of
blazars not in CGRaBS have been detected. This suggests that the CGRaBS (EGRET-like) blazar sample
is substantially different from that seen in the early \emph{Fermi} mission.This finding represents
a unique opportunity to investigate why gamma-ray activity is found only in certain blazars, and for
this reason we retain in our monitoring program all of the blazars in our original core sample even
if they have not yet been detected by the LAT. However, in order to optimize the potential for
studies of the cross-correlation between radio and gamma-ray light curves, we have since {\em added}
(and we continue to add) to our monitoring program all new \emph{Fermi}-LAT blazars north of
$-20\degr$ declination that are not CGRaBS members.

Several bright, stable non-blazar sources are included in our program to provide flux density
calibration and to monitor instrumental variability. These are 3C~48, 3C~161, 3C~286, DR~21, and
NGC~7027.  In addition to the stable sources, a number of bright sources are used to calibrate
pointing. These sources need not exhibit stable flux density, but need be brighter than about
100~mJy to permit pointing offsets to be measured reliably.

In addition to the core samples of blazars discussed above we have added further small samples of
objects to our bi-weekly monitoring program, including (i) any objects not already included in our
sample that are being studied in the F-GAMMA or VERITAS programs; (ii) a variety of galactic
objects, such as microquasars and cataclysmic variables; and (iii) a few bright radio galaxies that
show interesting jet properties.  We are continually adding sources of interest to our monitoring
sample, so that by now the sample comprises over 1500 objects that are monitored twice-weekly.

\subsection{Scheduling}

The large number of sources being observed requires the development of strategies to optimize the
use of the telescope and minimize the effect of known systematic errors. The principal systematic
errors we try to minimize are gain variations, atmospheric optical depth variations, and pointing
errors. To achieve this optimization while minimizing slew times and dead times between observations
requires careful planning.  Due to the size of our sample the scheduling must be automated.

Schedules are arranged to ensure that sources are observed between zenith angles of 20\degr\ and
60\degr\ whenever possible. This is done for a number of reasons:
\begin{enumerate}
\item the figure of the telescope was set for maximum gain in this elevation range;
\item at zenith angles less than about 20\degr\ the telescope has to move rapidly to track an object
and pointing accuracy can be degraded;
\item at zenith angles greater than about 60\degr\ ground spillover increases significantly with
decreasing elevation;
\item it is desirable to minimize the variation in atmospheric optical depth on our sources so as to
minimize this particular source of error; and
\item we try to minimize telescope slew times by observing to the south and east in a limited
elevation range.
\end{enumerate}

In the scheme we have developed, the sky is divided into 192 cells, each with a diameter $\la
20\degr$, using the HEALPix mesh with $N_{side}=4$~\citep{gorski_healpix:framework_2005}. Each
source is assigned to a cell.  From the sources in each cell, a pointing calibrator is selected
using the following criteria, applied in order: (1) if there is a flux calibrator in the region,
this source is selected; (2) if one or more sources in the region have a flux density larger than
500~mJy, the one which minimizes the average angular distance to all the sources in that region is
selected; (3) the source with the largest flux density in the region is selected.  For these flux
density comparisons, the median flux density of the source during the previous year's observations
is used.

Sources within the region are ordered to minimize slew time, using a direct search to find the
optimal order for regions with fewer than 9 sources and simulated annealing for regions with 9 or
more sources.  A second optimization step determines the order in which the regions are scheduled
using a heuristic algorithm in which regions are observed within a fixed zenith angle range and
regions to the south have priority. The total sample is observed in three days.

Prior to MJD 54906 (2009 March~16), a scheduling algorithm was used that did not enforce an angular
separation limit between pointing calibrators and subsequent flux density measurements.


\section{Data Editing and Calibration}

\subsection{Data Editing and Flagging}\label{sec:data_editing}

Editing and removal of corrupted data is performed using both automated and manual filters.

\subsubsection{Wind, Sun, Moon, and Zenith Angle Cuts}

Under high winds there is a systematic reduction in observed flux densities due to mis-pointing and
poor tracking.  Observations when the wind speed exceeds 6.7~$\mathrm{m\cdot s^{-1}}$ (15~mph) are
discarded.  To protect the telescope a ``wind watchdog'' program stows the telescope pointing at the
zenith when winds exceed steady $8.9~\mathrm{m\cdot s}^{-1}$ (20~mph) or gusts above
$13.4~\mathrm{m\cdot s}^{-1}$ (30~mph).  The telescope remains stowed until the wind speed has
remained below these thresholds for 1~hour.

Observations at zenith angles $<20\degr$ are discarded because the telescope is unable to track fast
enough in azimuth to match the sidereal rate near zenith.  The scheduling algorithm avoids
scheduling sources for observation at these zenith angles, so few observations are lost.
Observations at solar or lunar elongations less than 10\degr\ are also discarded.  The scheduler does
not avoid these areas of the sky so a small number of observations are lost.

\subsubsection{Pointing and Calibration Failures}
An observation is rejected if a pointing offset was not obtained within the prior 4800~s, or if the
pointing offset measurement immediately preceding the observation failed.  Occasional scheduling
errors resulted in observations without adequately measured pointing offsets.  These observations
are discarded.

An observation is rejected if fewer than two reliable calibration procedures using the CAL diode
were successfully executed within a two-hour interval centered on the time of the observation, or if
the difference between the largest and smallest CAL diode measurement within that interval differ by
more than 10\%.

\subsubsection{Saturation or Total Power Anomalies}
The total power varies depending on the attenuator setting, receiver gain fluctuations, atmospheric
conditions, and the observed zenith angle.  Observations that indicate saturation or other total
power anomalies are rejected. Heavy cloud cover or precipitation often causes large fluctuations in
total power. Such periods are identified by inspection of the total power time series and manually
discarded.  Negative flux density measurements are indicated by the 95\% upper limits on these
values.

\subsubsection{Measured Uncertainty}
We reject flux density measurements with anomalously large measured uncertainties, $\sigma_{15}$
(equation~(\ref{eq:flux_err})).  However, a straightforward cut at a fixed value or a fixed multiple
of the expected thermal uncertainty introduces a bias against larger flux densities.  This occurs
because there are contributions to the measured error that are proportional to the flux density of
the target radio source, such as telescope tracking errors.  We therefore apply a flux
density-dependent threshold and discard flux density measurements for which
\begin{equation}
  \label{eq:flux_sd_cut}
  \sigma_{15} > \zeta \sqrt{1 + \left(\rho\cdot S_{15}\right)^2}.
\end{equation}
The optimal values of $\zeta=0.0208$ and $\rho=0.2$ were estimated from the data to eliminate as many
unreliable measurements as possible while minimizing flux density bias.  About 2\% of the data is
eliminated by this filter.

\subsubsection{Switched Difference}\label{sec:swd}
We also use the switched difference $\mu$, defined by equation~(\ref{eq:swd}), to determine whether
flux density measurements might be contaminated by systematic errors.  The expected value of $\mu$
is 0, provided that the ground spillover and atmospheric noise in the \emph{ant} and \emph{ref}
beams are identical.  Pointing and tracking errors again give flux density-dependent contributions to
$\mu$, so to avoid bias against brighter radio sources we flag points where
\begin{equation}
  \label{eq:switched_diff_thresh}
  \left|\frac{\mu}{\sigma_{15}}\right| > \beta \cdot\frac{\left(\mu_{0}+\rho_{s}\cdot S_{15}\right)}{\sqrt{1 +\left(\rho_{t}\cdot S_{15}\right)^2}},
\end{equation}
Again the optimum values of the parameters ($\beta=5$, $\mu_{0}=1.148$, $\rho_{s}=0.0682$, and
$\rho_{t}=0.0243$) are determined from the data. This procedure gives consistent results across
calibration epochs and it discards about 2\% of flux densities, with comparable fractions dropped
from each epoch.

\subsection{Relative Calibration}\label{sec:relative_cal}
To correct for slow gain fluctuations of the receiver, we first divide each flux density measurement
by a calibration factor measured using the small noise diode \emph{cal}.  A measurement of the
strength of the \emph{cal} diode is made after each pointing observation, and no less than once per
hour.  Because gain fluctuations are slow, the calibration factor is averaged over a two-hour
window, centered on the time of the flux density measurement.  If there are fewer than two good
measurements of the strength of the \emph{cal} diode in that window then the flux density observation
is discarded.

Due to gravitational deformation of the telescope structure, the antenna gain varies substantially
with zenith angle.  We model this variation with a polynomial gain curve and scale flux density
measurements to remove the effect.  Additionally, the optimal axial focus position varies with
zenith angle, as well as solar zenith angle and elongation.  During observations, the focus position
is set using a polynomial model of the zenith angle variation and a correction is applied during
calibration using a more complete model that accounts for solar zenith angle and elongation.

The combined effect of these corrections is a factor, $\kappa_{rel}$, that is computed for each flux
density measurement.

\subsection{Absolute Calibration}\label{sec:absolute_cal}
We divide our observation period into epochs characterized by a consistent ratio between the
calibration diode and feed horn inputs to the receiver.  This ratio might change if, for example, the
signal path is disconnected and reconnected for maintenance, resulting in a slight change in loss
along one path.  Within a single epoch, the ratio of the calibration diode signal to a stable
astronomical source should therefore be constant.  Table~\ref{table:epochs} lists the epochs we have
used in our analysis.  Absolute calibration is applied to each epoch separately.
\begin{deluxetable}{ccccc}
  \tablecaption{Absolute calibration epochs. \label{table:epochs}}
  \tablehead{ \colhead{Epoch} & \colhead{MJD} & \colhead{Date} & \colhead{MJD} & \colhead{Date} }
  \tablewidth{0pt}
  \startdata
  1 & 54466 & 2008 Jan~01 & 54753 & 2008 Oct~14 \\
  2 & 54753 & 2008 Oct~14 & 54763 & 2008 Oct~24 \\
  3 & 54763 & 2008 Oct~24 & 55197 & 2010 Jan~01\\
  \enddata
\end{deluxetable}

For each epoch, a calibration factor is determined from regular observations of the primary
calibrator, 3C~286.  We adopt the spectral model and coefficients from~\citet{baars_absolute_1977}.
At our 15~GHz center frequency, this yields 3.44~Jy, with a quoted absolute uncertainty of about
5\%.  The calibration factor for epoch $i$, $\kappa_i$, is the ratio of the adopted flux density for the
calibrator to the weighted mean of the observations:
\begin{equation}
  \label{eq:abs_cal_factor}
  \kappa_i = {\mathrm{3.44\ Jy} \over {\left(\sum{S'_{15}\cdot{\sigma'_{15}}^{-2}}\right)/\left(\sum{{\sigma'_{15}}^{-2}}\right)}},
\end{equation}
where $S'_{15}$ and $\sigma'_{15}$ denote the flux densities for the calibrator with only the
relative calibration applied.

The total calibration factor for a flux density measurement in equation~(\ref{eq:flux}) is then
$\kappa=\kappa_{rel}\cdot\kappa_i\,$, and reflects both relative and absolute calibration.
Cross-checks of our calibration against 14.6~GHz observations of a number of common sources observed
with the Effelsberg 100~m telescope through the F-GAMMA project confirm the overall accuracy of our
flux density scale.

\subsection{Uncertainties in Individual Flux Density Measurements}\label{sec:main_unc}

In a perfect observing system with no sources of systematic error the uncertainties in the flux
density measurements would be given by the thermal noise on each observation.  In practice there are
many sources of systematic error, including the effects of weather and the atmosphere, mis-pointing
due to wind, and focus errors. Many of these are correctly identified and accounted for in the
automatic and manual editing described in \S~\ref{sec:data_editing}.  However, even after flux
density measurements affected by these problems are filtered out there remain many observations that
are significantly affected by systematic errors. Such systematic errors can lead to significant
errors in the measurement that are not reflected in the thermal noise of the observation and can
give rise to bad flux density measurements with small thermal errors.  This leads to ``outliers'' in
the light curves, i.e. points which do not lie close to the level determined from interpolation of
adjacent observations and which have small errors.  The task of identifying and eliminating or
allowing for the wide variety of systematic errors leading to such outliers is challenging and
time-consuming.  Great care must be taken not to assume that the behaviour of the source is known,
and hence to eliminate a real and potentially extremely interesting flux density variation.

We first apply an error model to determine the uncertainty of each flux density measurement:
\begin{equation}
  \label{eq:error_model}
  \sigma_{\rm total}^2 = \sigma_{15}^2 + \left(\epsilon\cdot S_{15}\right)^2 + \left(\eta\cdot \psi\right)^2\,,
\end{equation}
which is an extension of the model described in~\citet{angelakis_multi-frequency_2009}.  The first
term represents the measured scatter during the flux density measurement. This includes thermal
noise, rapid atmospheric fluctuations, and other random errors.  The second term adds an uncertainty
proportional to the flux density of the source.  This term allows for pointing and tracking errors,
variations in atmospheric opacity, and other effects that have a multiplicative effect on the
measured flux density.  In the third term $\psi$ is the switched power, defined by
equation~(\ref{eq:swp}). This term takes account of systematic effects that cause the A-B segment of
the flux density measurement to differ from the C-D segment, such as a pointing offset between the A
and D segments, or some rapidly varying weather conditions.

The error model is defined by the two parameters, $\epsilon$ and $\eta$, whose values must be
determined from the observations.  Because $\epsilon$ describes the error contribution due to
pointing errors, its value depends on whether a source is used as a pointing calibrator.
Furthermore, for non-pointing sources, $\epsilon$ is found to differ between the scheduling
algorithms used before and after MJD~54906 (2009 March~16).  The parameter, $\eta$, is found to be
adequately described by a single value for all sources and all epochs.  The adopted values are given
in Table~\ref{table:error_model_parameters}.

\begin{deluxetable}{cccc}
  \tablecaption{Error Model Parameter Values\label{table:error_model_parameters}}
  \tablehead{\colhead{Parameter} & \colhead{Pointing Calibrator} & \colhead{Early\tablenotemark{a}} & \colhead{Late\tablenotemark{a}}}
  \tablewidth{0pt}
  \startdata
  $\epsilon$ & 0.0057 & 0.0200 & 0.0135 \\
  $\eta$ & 3.173 & 3.173 & 3.173 \\
  \enddata
  \tablenotetext{a}{The ``early'' model applies prior to MJD 54906 (2009 March~16).}
\end{deluxetable}

For pointing sources, both $\epsilon$ and $\eta$ were estimated simultaneously using the stable flux
calibrators 3C~286, 3C~48, 3C~161, and DR~21.  Due to systematic errors, these sources and other
stable-flux density calibrators show long-term variations of 1--2\% so we fitted a 7th-order
polynomial to remove this trend from each source, then computed the residual standard deviation,
median flux density, the rms, and mean $\psi$ for each source, then used these to fit the error
model parameters.

\begin{figure}[t]
\includegraphics[width=\columnwidth]{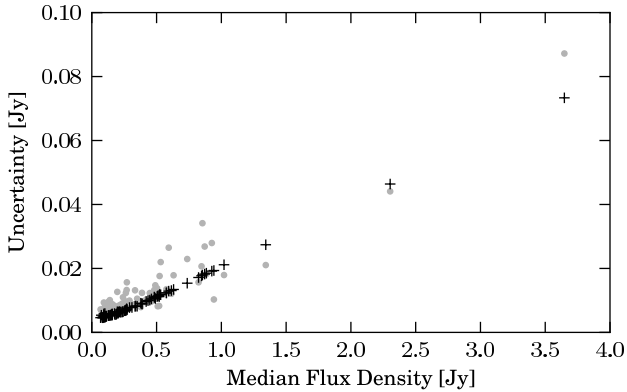}
\caption[early_epsilon_only_plot_detail]{\label{fig:error_model_early} Residual standard
  deviation (gray points) and fitted $\epsilon$-only error model values (black crosses) for ordinary
  sources in the early (MJD $<$ 54906) period.  The fit in the late period is similar.  A single
  high-flux density data point was omitted to limit the scale.}
\end{figure}

To determine the error model parameter $\epsilon$ for ordinary sources, we selected 100 sources that
exhibited little variation or slow, low-amplitude variations in flux density, between the start of
our program and MJD~55048 (2009 August~5).  This interval was split into two periods, ``early'' and
``late,'' at MJD~54906 and this procedure was separately applied to each period. For each light
curve, we fitted and removed a second-order polynomial trend, then iteratively removed outlier data
points with residuals greater than three standard deviations.  We repeated the fitting and outlier
removal until no further outliers were removed and we discarded any source with fewer than 10
remaining data points (retaining 94 and 88 sources in the early and late periods, respectively).
From the surviving points in each light curve, we computed the median and the rms flux densities,
and the standard deviation of the residuals.  We then fitted equation~(\ref{eq:error_model}) to these
data, omitting the $\eta$ term.  The data and the error model results for the early period are shown
in Figure~\ref{fig:error_model_early}.  We then adopted the same value of $\eta$ for these sources
as was determined for pointing sources.

\subsubsection{Long-Term Trends in 3C~286, 3C~274, and DR~21}
\begin{figure}[t]
\includegraphics[width=\columnwidth]{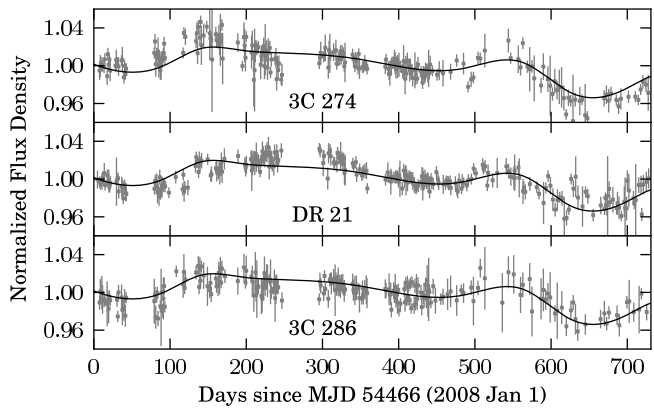}
\caption[prespline_cals]{\label{fig:prespline_cals} Normalized flux densities for 3C~274
  (top), DR~21 (center), and 3C~286 (bottom) after outlier removal.  Each light curve is normalized
  by its median.  The black line in each plot is the spline fit to the combined data.}
\end{figure}

After carrying out the above editing and calibration steps we returned to the residual 1--2\%
long-term ($\sim$ 6-month) variations in the light curves for stable-flux-density calibration
sources.  We chose 3C~286, 3C~274, and DR~21 for this study because they are well-known to be stable
on timescales of many years. The fractional variations in flux density of these objects are shown
in Figure~\ref{fig:prespline_cals} and are clearly correlated, indicating the presence of an
unidentified source of multiplicative systematic error. For each of these sources, we removed
2-$\sigma$ outliers in a 100-day sliding window and normalized the resulting data by the median flux
density.  We then combined the data for all three sources and fitted a cubic spline to the
result.

\begin{figure}[t]
\includegraphics[width=\columnwidth]{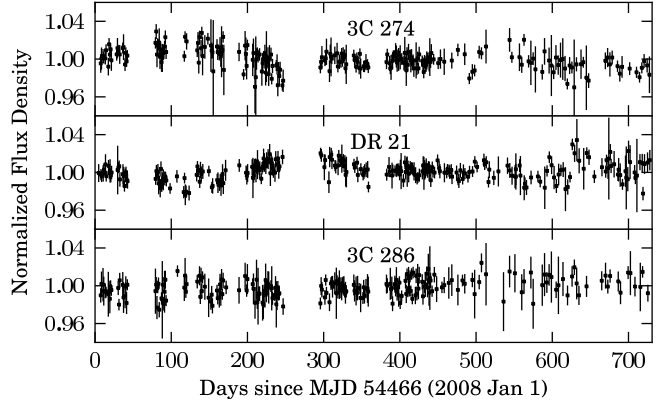}
\caption[postspline_cals]{\label{fig:postspline_cals} Normalized flux densities for 3C~274
  (top), DR~21 (center), and 3C~286 (bottom) after dividing by the spline fit to remove long-term
  systematic trends.}
\end{figure}

We apply the corresponding correction to all light curves in our program by dividing each flux
density by the value of this spline.  Figure~\ref{fig:postspline_cals} shows the residuals for the
three fitted sources after dividing out the spline fit.  The 1\% residual variation that remains is
the level of systematic uncertainty after correction for this long-term trend.


\subsection{Scaling of the Non-Thermal Error}

The reported error for each flux measurement has two qualitatively different components as described
in section \ref{sec:main_unc}. The first component is directly obtained during the flux measurement and
it represents random errors such as thermal noise and rapid atmospheric fluctuations, while the
second is introduced to take into account other, flux-density--dependent effects. This error model
requires the determination of constant factors, which we have called $\epsilon$ and $\eta$, and
which have been assumed to be source independent. However, many sources exhibit coherent long-term
variations with scatter about those clearly smaller than what would be expected as a result of the
quoted errors. This is a direct indication that in certain cases the simple assumption of
source-independent $\epsilon$ and $\eta$ resulted in overestimated errors.

\begin{figure}
  \begin{center}
    \includegraphics[width=\columnwidth]{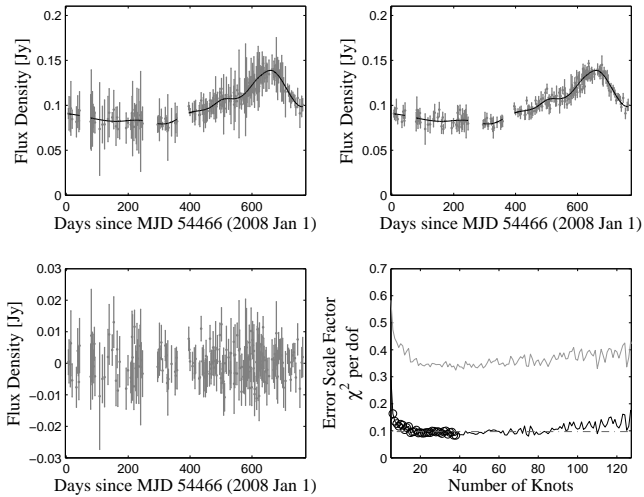}
    \caption[error_scale_factor_J0046+3900]{Example of the error scale factor correction using data for J0046+3900. The two upper
      panels show the light curve with the original (left) and corrected (right) error bars (gray
      points) and a typical spline fit (black line). The bottom left panel shows the residuals from
      the spline fit using the corrected error bars. In the bottom right panel, the $\chi^2$ per
      degrees of freedom (solid gray line) and correction factor (solid black line) are shown, with
      black circles marking the correction factors for fits that pass the acceptance tests, and a
      dashed line showing the adopted correction factor for the source.
      \label{ErrorScaleFactor}}
  \end{center}
\end{figure}

To correct these constant scale factors on a source-by-source basis, we have used cubic spline fits
and required a $\chi^2$ per degree of freedom to be one for the residuals. Due to the large number
of sources and the requirement of an uniform and consistent method for all the sources, an automatic
method was developed for this procedure.  For each source we can in principle use a range of number
of polynomial sections to construct a spline fit. We construct a spline fit for each possible number
of polynomial sections.\footnote{We use the MATLAB Spline Toolbox function spap2, which
  automatically selects the positions of the knots for the spline.} An outlier rejection filter
which uses a cubic spline fit with a small number of knots is used to fit the light curve. Points
with absolute residuals in the largest 5\% are not used for the following stage of the fitting
procedure. Not all the fits are acceptable, as some cases will have correlated residuals or a large
departure from normality. Acceptable fits are selected by using two statistical tests: Lilliefors
test for normality and the runs test for randomness.\footnote{We have used the implementation of
  both tests that are part of the MATLAB Statistics Toolbox.}  Only the fits for which both null
hypotheses cannot be rejected at the $10^{-3}$ level are considered acceptable. For each acceptable
fit, a scale factor that makes the $\chi^2$ per degree of freedom equal to one is calculated. Among
the scale factors for all the acceptable fits, the median scale factor is selected as the final
correction. The value of the scale factor is not very sensitive to the exact number of polynomial
sections. A typical example of the behavior of the scale factor is shown in Figure
\ref{ErrorScaleFactor}.

We have thus only rescaled the non-thermal part of the errors (the $S_{15}$ and $\psi$ terms in
equation~(\ref{eq:error_model})), and only for those sources for which the resulting correction factor
was smaller than one (i.e. the rescaling would result in {\em smaller} errors). The latter choice
was made for two reasons. First, a correction factor larger than one simply indicates that the
spline fit cannot provide an adequate description of the data.  This may result from a light curve
more variable than can be fit by spline with a given number of knots, so such a correction could
mask real variability.  Only the reverse is cause for concern---when the spline fit is too good a
fit, given the quoted errors. Second, this choice ensures a smooth transition between scaled and
non-scaled errors, as the transition point (correction factor equal to one) is equivalent to no
error scaling.

\section{Results and Analysis}

\begin{deluxetable*}{ccccccccc}
  \tablecaption{Program Source Data\tablenotemark{a}\label{TheTable}}
  \tablewidth{0pt}
  \tablehead{
    \colhead{Name} &
    \colhead{RA} &
    \colhead{Dec} &
    \colhead{z\tablenotemark{b}} &
    \colhead{Optical} &
    \colhead{Num\tablenotemark{c}} &
    \colhead{Flag\tablenotemark{d}} &
    \colhead{$S_{0}$\tablenotemark{e}} &
    {$\im$\tablenotemark{e}} \\
    &&&&\colhead{Class\tablenotemark{b}}&\colhead{Obs}&&\colhead{(Jy)}&(\%)}
  \startdata
  3C~48 & $01^{\rm h}37^{\rm m}41\fs30$ & \phs$33\degr09\arcmin35\farcs1$ & \nodata & \nodata & 281 & 0 & 1.805 $\pm$ 0.002 & $1.8\pm0.1$\\
  3C~161 & $06^{\rm h}27^{\rm m}10\fs12$ & $-05\degr53\arcmin05\farcs2$ & \nodata & \nodata & 191 & 0 & 2.081 $\pm$ 0.005 & $3.3\pm0.2$\\
  3C~274 & $12^{\rm h}30^{\rm m}49\fs42$ & \phs$12\degr23\arcmin28\farcs0$ & \nodata &  \nodata& 252 & 5 & 26.329 $\pm$ 0.021 & $0.9\pm0.1$\\
  3C~286 & $13^{\rm h}31^{\rm m}08\fs29$ & \phs$30\degr30\arcmin33\farcs0$ & \nodata & \nodata & 232 & 5 & 3.438 $\pm$ 0.003 & $0.7\pm0.1$\\
  DR~21 & $20^{\rm h}39^{\rm m}01\fs20$ & \phs$42\degr19\arcmin32\farcs9$ & \nodata & \nodata & 282 & 5 & 19.024 $\pm$ 0.011 & $0.7\pm0.1$\\
  J0001$+$1914 & $00^{\rm h}01^{\rm m}08\fs62$ & \phs$19\degr14\arcmin33\farcs8$ & 3.100 & FSRQ & 165 & 0 & 0.282 $\pm$ 0.003 & $11.5_{-0.7}^{+0.8}$\\
  J0001$-$1551 & $00^{\rm h}01^{\rm m}05\fs33$ & $-15\degr51\arcmin07\farcs1$ & 2.044 & FSRQ & 159 & 0 & 0.213 $\pm$ 0.002 & $8.9_{-0.6}^{+0.7}$  \\
  J0003$+$2129 & $00^{\rm h}03^{\rm m}19\fs35$ & \phs$21\degr29\arcmin44\farcs4$ & 0.450 & AGN & 169 & 0 & 0.087 $\pm$ 0.001 & $7.9_{-0.7}^{+0.8}$  \\
  J0004$+$2019 & $00^{\rm h}04^{\rm m}35\fs76$ & \phs$20\degr19\arcmin42\farcs2$ & 0.677 & BLL & 175 & 0 & 0.327 $\pm$0.003 & $12.5_{-0.7}^{+0.8}$ \\
  J0004$+$4615 & $00^{\rm h}04^{\rm m}16\fs13$ & \phs$46\degr15\arcmin18\farcs0$ & 1.810 & FSRQ & 154 & 0 & 0.181 $\pm$ 0.006 & $38.5_{-2.5}^{+2.8}$ \\
  J0004$-$1148 & $00^{\rm h}04^{\rm m}04\fs92$ & $-11\degr48\arcmin58\farcs4$ & \nodata & BLL & 106 & 0 & 0.720 $\pm$ 0.011 & $15.5_{-1.1}^{+1.2}$ \\
  \enddata
  \tablenotetext{a}{Only the first few rows are shown here; the complete version of this table is
    available in the electronic version of the journal.}

  \tablenotetext{b}{For the CGRaBS sample, redshift and optical classifications are repeated here
    from \citet{healey_cgrabs:all-sky_2008} for convenience.}

  \tablenotetext{c}{The number of observations that survived data editing and were used in our
    variability analysis.}

  \tablenotetext{d}{Variability analysis flag.  A value of 0 indicates a non-zero intrinsic
    modulation is found; 1 indicates the source is non-variable; 2 indicates insufficient
    observations for variability analysis; 3 indicates flux density too faint for variability
    analysis; 5 indicates the source was a calibrator used in the spline fit to remove long-term
    trends.}

  \tablenotetext{e}{Quoted errors are 1-$\sigma$ uncertainties. Values for non-variable sources
    indicate 3-$\sigma$ upper limits and no uncertainties are quoted for $\im$ or $S_0$.}
\end{deluxetable*}

\subsection{Monitoring Program Statistics}

Our target cadence was two flux density measurements per source per week, or about 200 measurements
per source in the first two years of the program.  Our resulting average effective cadence for
CGRaBS sources is about 134 measurements per source in the first two years of the program. The
efficiency compared to our nominal cadence is 67\%.

\subsection{Light Curves}
Light curves for the CGRaBS program sources are shown in Figures~7.1--7.1158.
Table~\ref{tab:fluxdensities} lists the filtered and calibrated 15~GHz flux density measurements
that result from the procedure described above.  Regular updates to the data set, including data for
sources outside the core sample released in this paper, are available from the program
website.\footnote{http://www.astro.caltech.edu/ovroblazars}

\begin{figure*}
  \begin{center}
    \includegraphics[width=\textwidth]{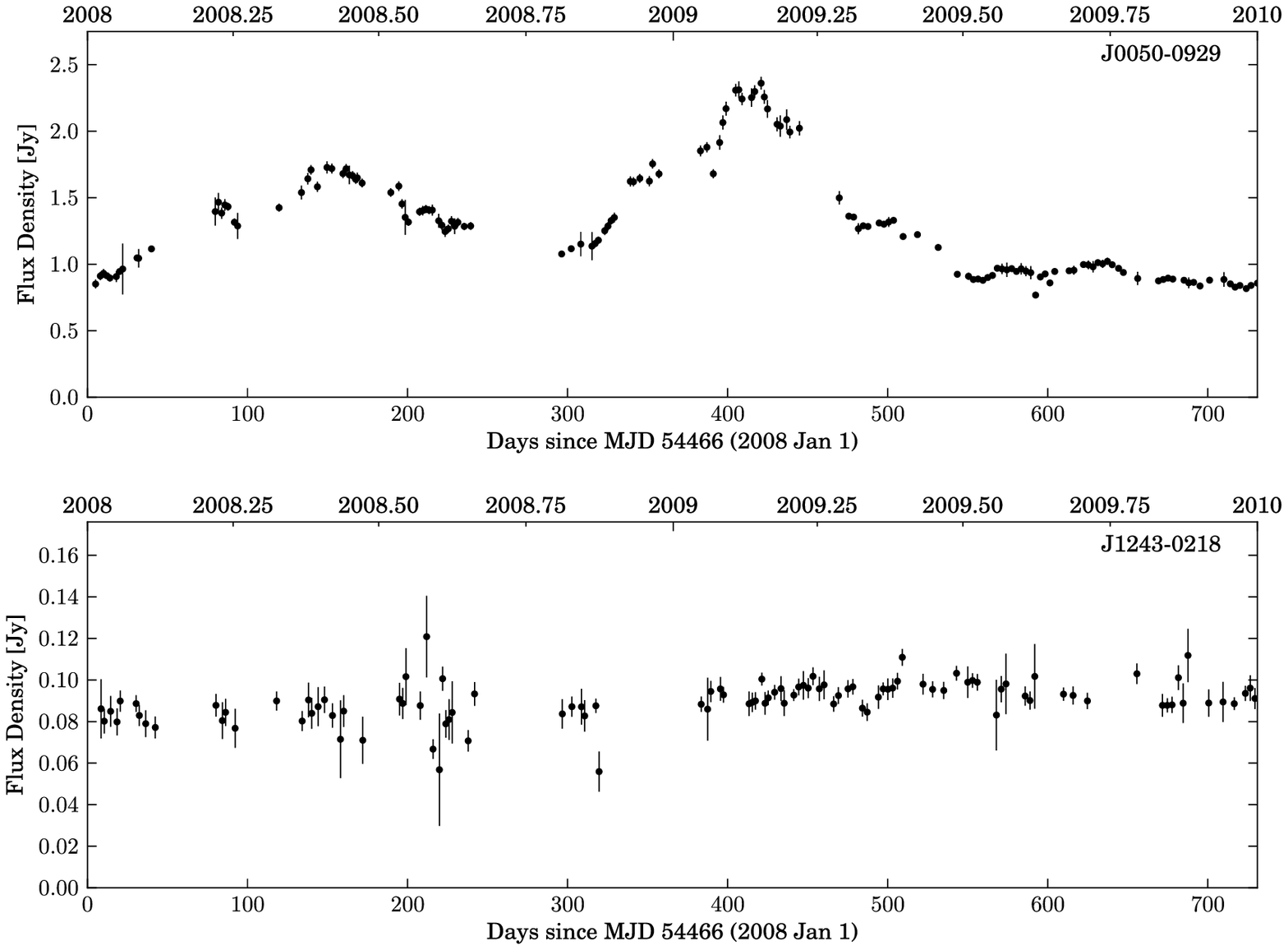}
    \caption[f7_print]{15~GHz light curves for calibrators and CGRaBS program sources.  Light curves for all sources
      are available in Figures~7.1--7.291 in the electronic version of the journal.
      \label{fig:lightcurves}}
  \end{center}
\end{figure*}

\begin{deluxetable}{ccc}
  \tablecaption{15~GHz Flux Densities\tablenotemark{a}\label{tab:fluxdensities}}
  \tablehead{\colhead{Source} & \colhead{MJD} & \colhead{Flux Density (Jy)}}
  \tablewidth{0pt}
  \startdata
  J0001$-$1551 & 54471.051377 & 0.244 $\pm$ 0.008 \\
  J0001$-$1551 & 54474.042836 & 0.232 $\pm$ 0.007 \\
  J0001$-$1551 & 54478.032303 & 0.221 $\pm$ 0.008 \\
  J0001$-$1551 & 54480.026840 & 0.238 $\pm$ 0.011 \\
  J0001$-$1551 & 54484.015903 & 0.229 $\pm$ 0.008 \\
  \enddata
  \tablenotetext{a}{Only the first few rows are shown here; the complete version of this table is
    available in the electronic version of the journal.}
\end{deluxetable}

\subsection{Source Variability}

In this section we discuss the variability {\em amplitude} observed in each source in our
sample. The variability properties of our sources in the time and frequency domains, as quantified
by measures such as the power spectrum and autocorrelation function, and correlation with gamma-ray
data to identify and measure time lags, will be discussed in detail in a forthcoming
publication~\citep{max-moerbeck_ovro_2010}.

The questions of the variability amplitude of a source and the confidence with which this can be
measured are complex ones and have been traditionally addressed using a variety of measures and
tests, such as the variability index \citep[e.g.,][]{aller_pearson-readhead_1992}; the fluctuation
index \citep[e.g.,][]{aller_pearson-readhead_2003}; the modulation index
\citep[e.g.,][]{kraus_intraday_2003}; the fractional variability amplitude
\citep[e.g.,][]{edelson02,soldi08}; and $\chi^{2}$ tests of a null hypothesis of
non-variability. Each of these tools provides different insights to the variability properties of
sources and is sensitive to different uncertainties, biases, and systematic errors.  For example,
the variability index, defined as the peak-to-trough amplitude change of the flux, is a measure of
the amplitude of the variability of a source:
\begin{equation}
  V = \frac{(S_{\rm max} - \sigma_{\rm max})-(S_{\rm min}+\sigma_{\rm min})}
  {(S_{\rm max}-\sigma_{\rm max}) +(S_{\rm min} + \sigma_{\rm min})}\,,
\end{equation}
where $S_{\rm max}$ and $S_{\rm min}$ are the highest and lowest measured flux densities,
respectively, and $\sigma_{\rm max}$ and $\sigma_{\rm min}$ are the uncertainties in these
measurements. Although the definition is constructed to account for the effect of measurement
uncertainties, the quantity is well-defined only when variability is significantly greater than
measurement errors, and it can yield negative values for sources with low signal-to-noise ratios or
little intrinsic variability.  In addition, it is very sensitive to outliers and is not robust
against random Gaussian excursions from the mean. Such excursions are to be expected for sources
that are regularly monitored over long periods of time: even non-variable sources are likely to have
at least one pair of 2-$\sigma$ high and low measurements after being observed more than 100 times,
as is the case for most sources in our sample.

An associated measure of variability amplitude is the modulation index, defined as the standard
deviation of the flux density measurements in units of the mean measured flux density,
\begin{equation}
m_{\rm data} = \frac{\sqrt{\frac{1}{N}\sum_{i=1}^N\left(S_i- \frac{1}{N}\sum_{i=1}^NS_i
\right)^2}}{\frac{1}{N}\sum_{i=1}^NS_i}\,.
\label{fromdata}
\end{equation}
The modulation index has the advantage that it is always non-negative and more robust
against outliers. However, it still represents a convolution of intrinsic source variation and
observational uncertainties: a large modulation index could be indicative of either a strongly
variable source or a faint source with high uncertainties in individual flux density measurements. For this reason,
the correct interpretation of results on the modulation index requires that measurement errors and
the uncertainty in $m$ due to the finite number of flux density measurements be properly accounted for.

One method that has been widely used to evaluate the information encoded in variability measures is
to evaluate each measure for a set of constant-flux-density calibrators, which are known to have a flux density
constant in time and have been observed with the same instrument over the same periods of time. The
value of the variability measure obtained for the calibrators is then used as a threshold value, so
that any source with variability measure equal to or lower than that of the calibrators is
considered consistent with being non-variable. However, a variability measure value higher than that
of the calibrators is a necessary but not sufficient condition for establishing variability:
calibrators are generally bright sources, with relative flux density measurement uncertainties typically
lower than the majority of monitored sources; additionally, variability measures are affected by the
sampling frequency, which is not necessarily the same for all monitored sources and the calibrators.

Alternatively, the {\em significance} of variability in a given source can be established through
tests (such as a $\chi^{2}$ test) evaluating the consistency of the obtained set of measurements
with the hypothesis that the source was constant over the observation interval. However, such tests
provide very little information on sources for which statistically significant variability cannot be
established, as they cannot distinguish between {\em intrinsically non-variable sources} and sources
that could be {\em intrinsically variable} but inadequately observed for their variability to be
revealed.

Here we propose a new index for characterizing source variability: the {\em intrinsic modulation
  index} $\im$, which is the intrinsic standard deviation of the distribution of source flux densities in
time, $\sigma_0$, measured in units of the intrinsic source mean flux density, $S_0$. Here the term
``intrinsic'' is used to denote flux densities and variations as would be observed with perfectly uniform
sampling of adequate cadence and zero observational error:
\begin{equation}
\im = \frac{\sigma_0}{S_0}\,.
\end{equation}
In this way, $\im$ is a measure of the true amplitude of variations in the source, rather than a
convolution of true variability, observational uncertainties, and effects of finite
sampling. Observational uncertainties and finite sampling will, of course, affect the accuracy with
which $\im$ can be measured. The purpose of the analysis described in this section is to derive a
best estimate of $\im$, as well as an estimate of the uncertainty in our measurement of this
quantity. For sources with $\im$ within $3\sigma$ from zero, the $3\sigma$ upper limit on $\im$ will
be evaluated.

\subsubsection{A Likelihood Analysis to Obtain the Intrinsic Modulation Index}

For the purposes of our analysis, we will assume that the ``true'' flux densities for each AGN are
normally distributed, with mean $S_0$, standard deviation $\sigma_0$, and intrinsic modulation index
$\im=\sigma_0/S_0$. We have $N$ measurements of the flux density, $S_j$, each of which has an
associated observational uncertainty, also assumed Gaussian, $\sigma_j$.

Let us assume that at a moment of observation, a source has a ``true'' flux density $S_t$. The
probability density to observe a value near $S_j$ if the observational uncertainty is $\sigma_j$ is
\begin{equation}
p(S_t, S_j, \sigma_j) = \frac{1}{\sigma_j \sqrt{2\pi}} \exp \left[
-\frac{(S_t-S_j)^2}{2\sigma_j^2}\right]\,.
\end{equation}
In addition, the probability density that the true source flux density at one of the moments of
observation is near $S_t$ if the source flux densities are distributed normally with mean $S_0$ and
standard deviation $\sigma_0$ is
\begin{equation}
p(S_t, S_0, \sigma_0) = \frac{1}{\sigma_0 \sqrt{2\pi}} \exp \left[
-\frac{(S_t-S_0)^2}{2\sigma_0^2}\right]\,.
\end{equation}
Therefore, the likelihood of observing one flux density $S_j$ with uncertainty $\sigma_j$ from the
particular source is
\begin{equation}\label{oneint}
\ell_j = \int_{\rm all \,\, S_t} \!\!\!\!\!\!\!\!dS_t \frac{\exp \left[
-\frac{(S_t-S_j)^2}{2\sigma_j^2}\right] }{\sigma_j \sqrt{2\pi}} \frac{\exp \left[
-\frac{(S_t-S_0)^2}{2\sigma_0^2}\right]}{\sigma_o \sqrt{2\pi}} \,,
\end{equation}
which amounts to calculating the probability to observe $S_j$ through any possible true flux density value
$S_t$.  If the limits of integration above are taken to be from $S_t = -\infty$ to $S_t=\infty$ then
the integral has an analytic solution~\citep[see, e.g.,][]{venters_pavlidou_2007}:
\begin{equation}
\ell_j = \frac{1}{\sqrt{2\pi(\sigma_0^2+\sigma_j^2)}}\exp \left[
-\frac{(S_j-S_0)^2}{2(\sigma_j^2+\sigma_0^2)}\right]\,.
\end{equation}

The likelihood for $N$ observations ($S_j,\sigma_j$) for $j=1,...N$ is
\begin{eqnarray}
\mathcal{L}(S_0,\sigma_0) = \prod_{j=1}^N\ell_j &=&  \left( \prod_{j=1}^N
\frac{1}{\sqrt{2\pi(\sigma_0^2 + \sigma_j^2)}}\right)\times \nonumber \\
&&  \exp \left[
-\frac{1}{2}\sum_{j=1}^N\frac{(S_j-S_0)^2}{\sigma_j^2 +\sigma_0^2}\right]\,.
\end{eqnarray}
The intrinsic standard deviation $\sigma_0$ can be eliminated in favor of the intrinsic modulation
index,
\begin{equation}
\sigma_0= \im S_0,
\end{equation}
so that
\begin{eqnarray}\label{joint}
\mathcal{L}(S_0,\im) &=& S_0\left( \prod_{j=1}^N \frac{1}{\sqrt{2\pi(\im^2S_0^2 +
\sigma_j^2)}}\right) \times \nonumber \\ && \exp \left[ -\frac{1}{2}\sum_{j=1}^N\frac{(S_j-S_0)^2}{\sigma_j^2
+\im^2S_0^2}\right]
\end{eqnarray}
This likelihood is symmetric about $\im=0$, as $\im$ only enters through its square. For
this reason, this formalism can  guarantee non-negative intrinsic modulation indices without loss of information.

\begin{figure}[t]
\includegraphics[width=\columnwidth, clip=]{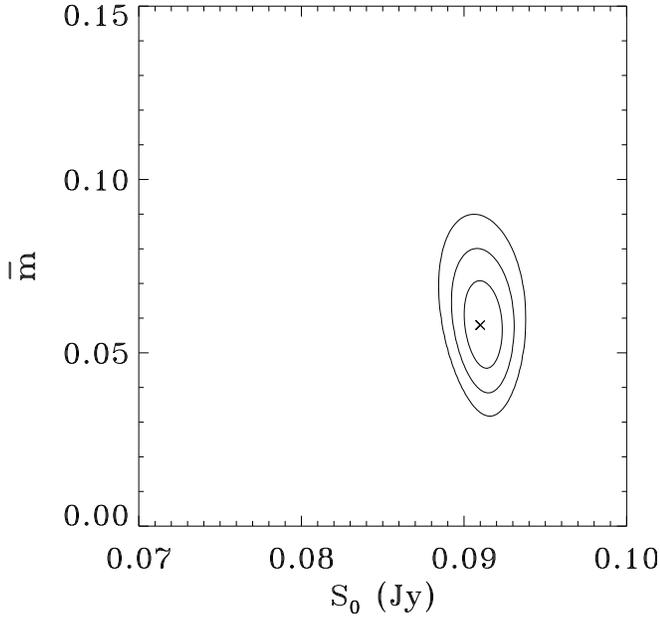}
\caption[contsJ1243-0218]{\label{contours} 1, 2, and $3\sigma$ contours of the joint likelihood
  $\mathcal{L}(S_0,\im)$ for blazar J1243$-$0218. }
\end{figure}

\begin{figure}[t]
\includegraphics[width=\columnwidth, clip=]{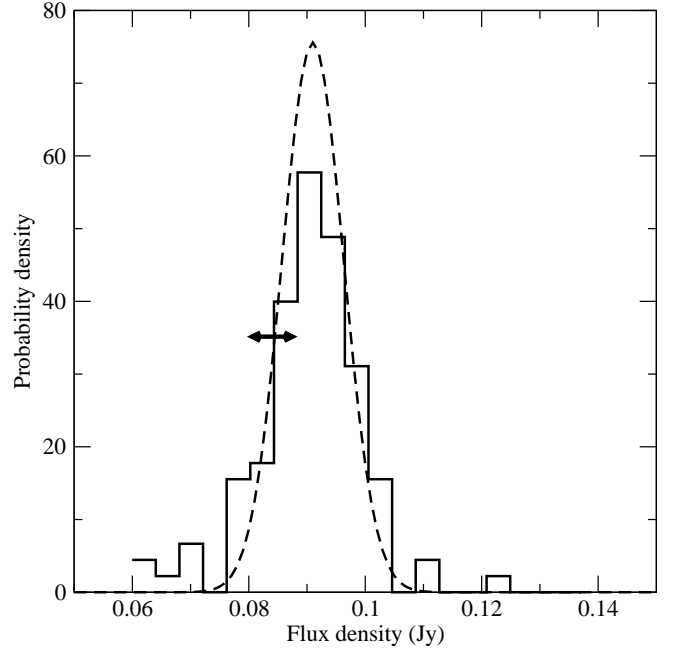}
\caption[histogram_fluxes]{\label{maxLmodel}  Maximum-likelihood Gaussian model for the flux density distribution (dashed
line), plotted over the histogram of measured flux densities (solid line) for blazar J1243$-$0218. The
arrow indicates the size of the typical measurement uncertainty. }
\end{figure}

Maximizing the joint likelihood $\mathcal{L}(S_0,\im)$, we can derive maximum-likelihood estimates
of $S_0$ and $m$. Isolikelihood contours containing 68.26\%, 95.45\%, and 99.73\% of the total
volume under the joint likelihood surface define the 1, 2, and $3\sigma$ contours respectively (see
Figure~\ref{contours} for an example in the case of J1243$-$0218, whose light curve is shown in
Figure~\ref{fig:lightcurves}). The maximum-likelihood Gaussian for the distribution of flux
densities for the same object is compared to the histogram of measurements in
Figure~\ref{maxLmodel}. Note that the maximum-likelihood Gaussian is narrower than the histogram;
this behavior is expected, as the histogram is a representation of measurements sampling the
underlying distribution with finite error. The typical magnitude of the latter for the particular
source is shown in Figure~\ref{maxLmodel} with the blue arrows, and it is indeed comparable with the
difference in width between the maximum-likelihood Gaussian and the histogram.

\begin{figure}[t]
\includegraphics[width=\columnwidth, clip=]{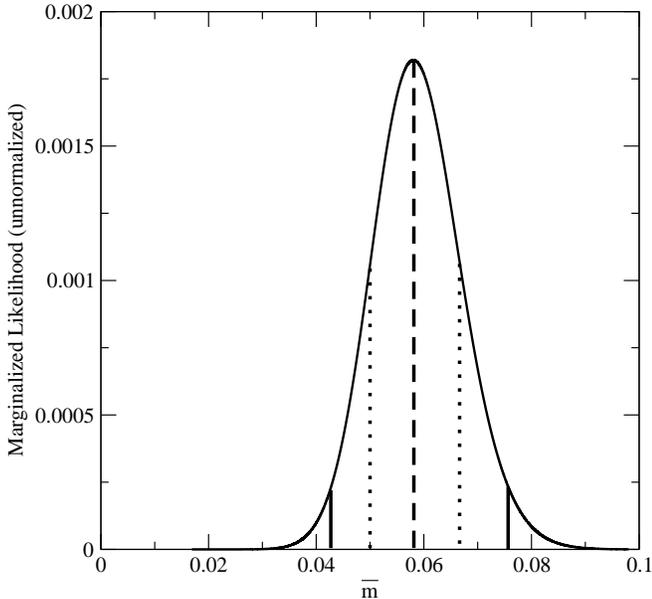}
\caption[marginalizedforplot]{\label{marginalized}  Marginalized likelihood $\mathcal{L}(m)$ for J1243$-$0218 (solid
curve). Dashed vertical line: best-estimate $\im$; dotted vertical lines: $1\sigma$ $\im$ range; solid vertical lines:
$2\sigma$ $\im$ range.}
\end{figure}

To derive the most likely value of $\im$ and the associated uncertainties {\em regardless} of the
true value of $S_0$, we integrate $S_0$ out of $\mathcal{L}(S_0,\im)$, and obtain the marginalized
likelihood as a function of only $\im$:
\begin{eqnarray}
\mathcal{L}(\im) =\int _{\rm all  \,\, S_0}\!\!\!\!\!\!&dS_0& S_0 \left\{\left( \prod_{j=1}^N
\frac{1}{\sqrt{2\pi(m^2S_0^2 + \sigma_j^2)}}\right) \right. \times \nonumber \\ && \left. \exp \left[
-\frac{1}{2}\sum_{j=1}^N\frac{(S_j-S_0)^2}{\sigma_j^2 +m^2S_0^2}\right]\right\}\,.
\end{eqnarray}
Then, the value of $\im$ that maximizes the marginalized likelihood is our best estimate of it, and
the 1-$\sigma$ uncertainty on the modulation index can be found by locating the isolikelihood
$\im-$values $\im_1$ and $\im_2$ for which
\begin{equation}
 \mathcal{L}(\im_1) = \mathcal{L}(\im_2)
\end{equation}
and
\begin{equation}\label{uncs}
\frac{\int_{\im_1}^{\im_2}\mathcal{L}(\im) d\im
}{\int_0^{\infty}\mathcal{L}(\im) d\im } = 0.6826\,.
\end{equation}
Note that the upper and lower $1\sigma$ errors are not generally symmetric in our formalism.
Similarly, $2\sigma$ and $3\sigma$ ranges can be derived by substituting the right-hand-side of
equation~(\ref{uncs}) by $0.9545$ and $0.9973$, respectively.  The marginalized likelihood,
best-estimate $\im$, and the $1\sigma$ and $2\sigma$ $\im$ ranges for blazar J1243$-$0218 are shown
in Figure~\ref{marginalized}.

If the maximum-likelihood $\im$ is less than $3\sigma$ away from $\im=0$, we consider that
statistically significant variability cannot be established. In these cases, we calculate the
$3\sigma$ upper limit on $\im$, which is defined as the value $\im_3$ for which
\begin{equation}\label{uplim}
\frac{\int_0^{\im_3}\mathcal{L}(\im) d\im }{\int_0^{\infty}\mathcal{L}(\im) d\im
} = 0.9973\,.
\end{equation}

The use of the intrinsic modulation index and the likelihood analysis we have employed to estimate
it have the advantage of offering a way to obtain information about the {\em intrinsic} variability of the
source, deconvolved from observational errors in individual flux density measurements and the effects of finite
sampling, while providing strictly defined $1,2$ and $3\sigma$ uncertainties for our estimate of $\im$
(essential when conducting population studies), and upper limits for $\im$ when variability cannot be
established at a $\ge 3\sigma$ confidence. However, our choice carries its own caveats.

(1) {\em Model-dependence of the likelihood analysis.} A functional form has to be assumed for the
intrinsic distribution of flux densities (here we have assumed it to be Gaussian), resulting in a loss of
generality.  The validity of this assumption can be tested by comparing the maximum-likelihood
intrinsic flux density distribution to the histogram of measured flux densities, to evaluate whether the
maximum-likelihood flux density distribution is a reasonable description of the data (modulo observational
uncertainties). This is indeed the case for many, although not all, of our sources. An example of a
source well described by the maximum-likelihood flux density distribution is shown in
Figure~\ref{maxLmodel}. Other sources however show bimodality, and the distribution of measured
flux densities in these cases could be better described by, for example, a double Gaussian. An extended
likelihood analysis that does explicitly account for bimodality and calculates not only the
intrinsic modulation index but also duty cycles and flaring-to-quiescent flux density ratios will be
presented in an upcoming publication. For the purposes of this work we have confirmed that, even
when a single Gaussian is not an adequate description of the flux density distribution, the intrinsic
modulation $\im$ index is well-correlated, within uncertainties, with the modulation index $m_{\rm
  data}$ derived from the data (equation~(\ref{fromdata})), which, although contaminated by
observational uncertainties, is completely model-independent (see next section).

(2) {\em Assumption of unbiased sampling.} Our analysis assumed that the flux density values we have not
sampled are not correlated with each other.  This assumption is poor in the case of lengthy outages,
as well as for increased cadence for any single epoch. In our analysis we have disregarded the
additional data taken during epochs of increased cadence for specific objects (during, for example,
campaigns to constrain intra-day variability).

(3) {\em Leakage of probability density to negative flux densities.} In certain cases, extending the
integration over intrinsic flux densities from $-\infty$ to $\infty$ (in equation~(\ref{oneint})) to
simplify the mathematical manipulations leads to unacceptable leakage of probability density to
unphysical domain of negative true flux densities. This approximation is adopted for numerical
efficiency, since in this case the likelihood can be expressed analytically without the need to
perform multi-dimensional integrals for every object in our large sample. For most objects in our
sample the leakage to the negative flux density domain is negligible. The error introduced in this
way becomes important only for very dim AGN (because the peak of the $S_t$ distribution is very
close to zero) or very variable AGN (because of very long tails in the $S_t$ distribution). None of
the CGRaBS sources in our sample are dim enough for the first effect to be a problem, and very few
are variable enough: for $\im \sim 0.5$, about $3\%$ of the ``true flux density'' probability
density leaks to negative values, with the problem becoming more severe for more variable sources;
only four CGRaBS sources have $\im \ge 0.5$.

\subsubsection{Variability Analysis---Results}\label{vanalysis}

\begin{figure}[t]
\includegraphics[width=\columnwidth]{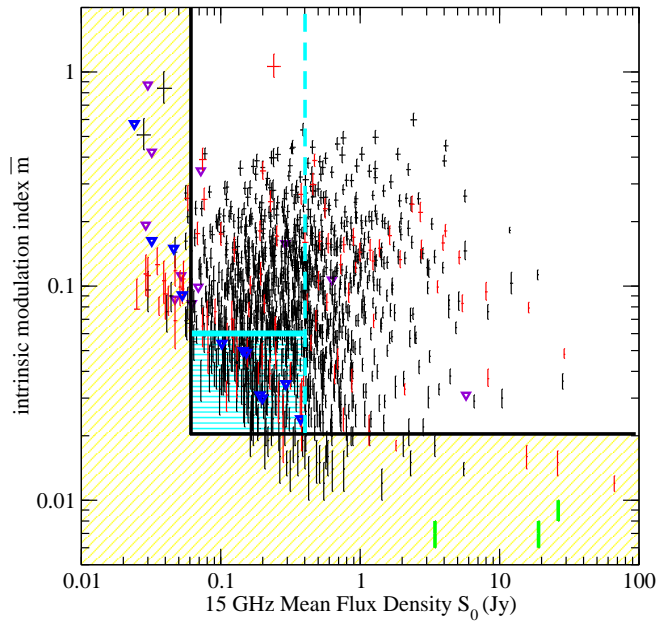}
\caption[mScatterPlot]{\label{all_m_vs_s} Intrinsic modulation index $\im$ and associated 1-$\sigma$ uncertainty,
  plotted against intrinsic maximum-likelihood average flux density, $S_0$, for all sources in the
  OVRO monitoring program which have enough (more than 3) acceptable, non-negative flux density
  measurements. Black points: CGRaBS sources; red points: non-CGRaBS sources; green points:
  calibrators; blue triangles: 3-$\sigma$ upper limits for CGRaBS sources for which variability
  could not be established at $\ge 3\sigma$ confidence level; magenta triangles: as blue triangles, for
  non-CGRaBS sources. The error bar on $S_0$ corresponds to the 1-$\sigma$ uncertainty in mean flux density,
  calculated from the joint likelihood (equation~(\ref{joint})) marginalized over $\im$. Variable
  CGRaBS sources outside the yellow and cyan shaded areas are used in the population studies of
  \S~\ref{popstudies}.}
\end{figure}

In Figure~\ref{all_m_vs_s} we plot the intrinsic modulation index $\im$ and associated 1-$\sigma$
uncertainty against the intrinsic, maximum-likelihood average flux density, $S_0$, for all sources in the
OVRO monitoring program (CGRaBS and non-CGRaBS sources).  The error bar on $S_0$ corresponds to the
1-$\sigma$ uncertainty in mean flux density, calculated from the joint likelihood (equation~(\ref{joint}))
marginalized over $\im$. CGRaBS sources are shown in black, while non-CGRaBS sources are shown in
red.

Variability could only be established at the 3-$\sigma$ confidence level or higher for 1146 out of
1158 CGRaBS blazars in our sample.  For this study, we considered only sources for which at least
3~flux densities were measured, a positive mean flux density $\ge 2\sigma$ from zero was found, and
at least 90\% of the individual flux density measurements were $\ge 2\sigma$ from zero.  These
criteria excluded one source.  For the other 11 sources we have calculated 3-$\sigma$ upper limits
for $\im$.  We plot these upper limits with blue triangles. We also plot, with magenta triangles,
3-$\sigma$ upper limits for non-CGRaBS sources that were consistent with non-variable and had enough
flux density measurements.

Calibration sources 3C~286, DR~21, and 3C~274 are shown in green. Although these sources
are the least variable (as expected) of all sources in which variability can be established and a
non-zero $\im$ can be measured, $\im$ for these sources is finite and measurable. This means that
some residual long-term variability remains in our calibrators beyond what can be justified by
statistical errors alone. This could conceivably result from true calibrator source variation, but
more likely reflects incomplete removal of small-amplitude calibration trends.  Because $\im< 1\%$
for these three sources, we quote a systematic uncertainty $\Delta \im_{\rm syst} = 0.01$ for the
values of the intrinsic modulation index we produce through our analysis.

To ensure that our population studies are not affected by this residual systematic variability, in
all analyses discussed in \S~\ref{popstudies} only sources with $\im \ge 0.02$ will be used, so that
we remain comfortably above this $1\%$ systematic uncertainty limit. In addition, for sources with
$S_0\le 60~\mathrm{mJy}$, the number of sources for which variability can be established is of the same order
as the number of sources (both CGRaBS and non-CGRaBS) for which we could only measure an upper
limit, and these upper limits are very weak and non-constraining. For this reason, we also exclude
from our population studies of \S~\ref{popstudies} all sources below $S_0=60~\mathrm{mJy}$. The part of the
parameter space excluded due to these two criteria is shown in Figure~\ref{all_m_vs_s} as the yellow
shaded area bounded by the solid black lines.

For $S_0\ge 0.4~\mathrm{Jy}$, no obvious correlation between flux density and modulation index is apparent,
and no CGRaBS sources exist with upper limits above our cut of $\im = 2\%$. However, for sources
with $S_0 < 0.4~\mathrm{Jy}$, there is an absence of points in the lower-left corner of the allowed
parameter space defined by the thick solid lines: for faint sources, we can only confidently
establish variability if that variability is strong enough. The effect disappears for variability
amplitudes greater than about $6\%$. In addition, there are no CGRaBS upper limits higher than $6\%$
for sources brighter than $60~\mathrm{mJy}$. We conclude that we are able to measure variability at
the level of $6\%$ or higher for {\em any} CGRaBS source brighter than $60~\mathrm{mJy}$.

To ensure that our population studies are not affected by our decreased efficiency in measuring
variability in sources with $60~\mathrm{mJy}\le S_0 \le 0.4~\mathrm{Jy}$ and $2\%\le\im \le 6\%$, we
will also exclude this part of the $(S_0,\im)$ parameter space from our analysis in
\S~\ref{popstudies}. The part of the parameter space excluded due to these criteria is shown in
Figure~\ref{all_m_vs_s} as the cyan shaded area.

The single point with $\im \geq 1.0$ near the upper left corner of the plot is Cygnus X-3, a
galactic microquasar not part of our core program, which indeed exhibits a strong flare during which
the flux density increases by about an order-of-magnitude over its average in a short period of
time. Although in this case our underlying assumption of the flux density measurements being
consistent with a single Gaussian distribution is clearly not a sufficient description of the data
(rather, this case is much better described by a bimodal distribution), the qualitative result of
variability with very high amplitude in this source is robust.

\begin{figure}[t]
  \includegraphics[width=\columnwidth]{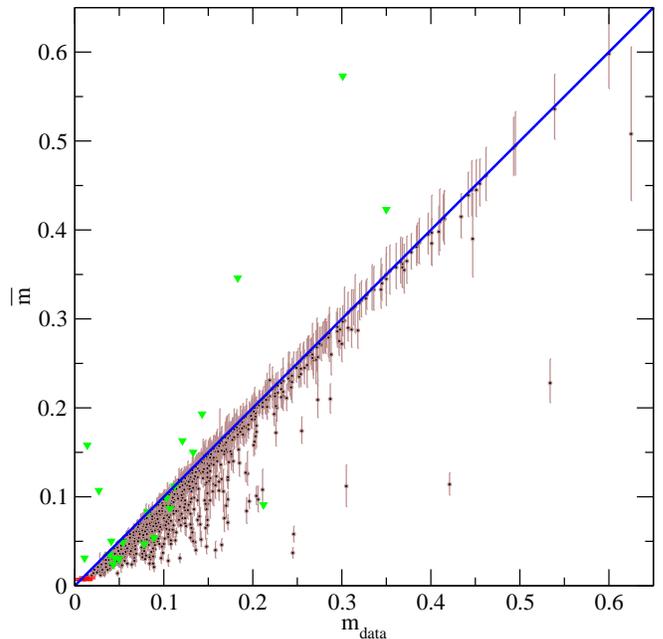}
  \caption[m_vs_mdata]{\label{all_m_vs_m} Intrinsic modulation index $\im$ and associated 1-$\sigma$
    uncertainty, plotted against the ``raw'' modulation index, $m_{\rm data}$, of
    equation~(\ref{fromdata}) as black points with brown error bars. The $\im=m_{\rm data}$ line is
    shown in blue. Green triangles are the 3-$\sigma$ upper limits of sources for which variability
    could not be established. Calibrators are plotted in red. }
\end{figure}

In Figure~\ref{all_m_vs_m} we plot the intrinsic modulation index $\im$ and associated 1-$\sigma$
uncertainty against the ``raw'' modulation index $m_{\rm data}$ of equation~(\ref{fromdata}). The
$\im=m_{\rm data}$ line is shown in blue. Green triangles are the 3-$\sigma$ upper limits of sources
for which variability could not be established. Calibrators 3C~286, DR~21, and 3C~274 are plotted in
red. Since apparent variability due to the finite accuracy with which individual flux densities can
be measured has been corrected out of $\im$, the expectation is that deviations from the $\im=m_{\rm
  data}$ will be more pronounced for sources that are not intrinsically very variable (so that the
scatter in the flux density measurements is appreciably affected, and even dominated, by measurement
error). In addition, deviations are expected to be below the line, as $\im$ should be {\em smaller}
than $m_{\rm data}$. Both these expectations are verified by Figure~\ref{all_m_vs_s}.  Note that
upper limits need not satisfy this criterion, as the ``true'' value of the modulation index can take
any value below the limit. Upper limits above the blue line are weak, indicating that the reason
variability could not be established is the poor sampling or quality of the data, and not
necessarily a low intrinsic variation in the source flux density.

\begin{figure}
\includegraphics[width=\columnwidth, clip=]{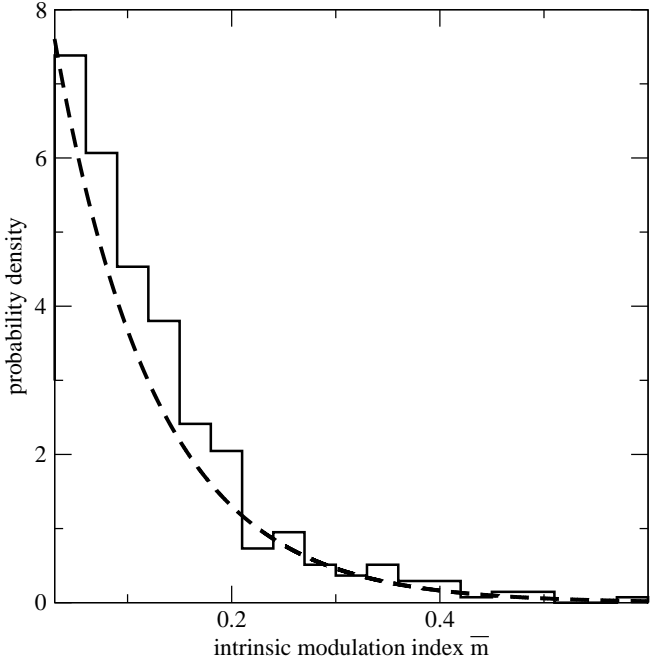}
  \caption[pop_histogram]{Histogram of maximum-likelihood intrinsic modulation indices
    $\im$, for the 456 CGRaBS blazars with $S_0>400~\mathrm{mJy}$. The dashed line represents an
    exponential distribution with $\langle \im \rangle = 0.091$.}
  \label{histdist}
\end{figure}

For the 456 CGRaBS objects which have $S_0>400~\mathrm{mJy}$ and for which variability can be
established, we plot, in Figure~\ref{histdist}, a histogram of their intrinsic modulation indices
$\im$ normalized so that the vertical axis has units of probability density. The dashed line
represents an exponential distribution of mean $\langle \im \rangle = 0.091$ which, as we can see,
is an excellent description of the data. Motivated by this plot, we will be using the monoparametric
exponential family of distributions:

\begin{equation}\label{expdist}
f(m)dm = \frac{1}{m_0} \exp\left[-\frac{m}{m_0}\right]dm
\end{equation}
with mean $m_0$  and variance $m_0^2$, to characterize various sub-samples of our blazar sample.

\subsubsection{Variability Analysis---Population Studies---Formalism}\label{popstudies}

We now turn our attention to whether the intrinsic variability amplitude at 15~GHz, as quantified by
$\im$, correlates with the physical properties of the sources in our sample. To this end, we will
determine the {\em distribution} of intrinsic variability indices $\im$ for various subsets of our
monitoring sample, and we will examine whether the various subsets are consistent with being drawn
from the same distribution.

We will do so using again a likelihood analysis. We will assume that the distribution of $\im$ in
any subset is an exponential distribution of the form given in equation~(\ref{expdist}). Since
distributions of this family are uniquely described by the value of the mean, $m_0$, our aim is to
determine $m_0$, or rather the probability distribution of possible $m_0$ values, in any specific
subset.

The likelihood of a single observation of a modulation index $\im_i$ of Gaussian uncertainty
$\sigma_i$ drawn from an exponential distribution of mean $m_0$ is
\begin{eqnarray}
\ell_i &=& \int_{\im=0}^{\infty} d\im \frac{1}{m_0} \exp\left(-\frac{\im}{m_0}\right)
\frac{1}{\sigma_i\sqrt{2\pi}} \exp\left[-\frac{(\im-\im_i)^2}{2\sigma_i^2}\right]
\nonumber \\
&=& \frac{1}{m_0\sigma_i\sqrt{2\pi}}
\exp\left[-\frac{\im_i}{m_0}\left(1-\frac{\sigma_i^2}{2m_0\im_i}\right)\right]\times \nonumber \\
&&\int_{\im=0}^{\infty} d\im \exp\left[-\frac{[\im-(\im_i-\sigma_i^2/m_0)]^2}{2\sigma_i^2}\right]\,,
\end{eqnarray}
where, to obtain the second expression, we have completed the square in the exponent of the
integrand. The last integral can be calculated analytically, yielding
\begin{eqnarray}
\nonumber \\
\ell_i&=& \frac{1}{2m_0}
\exp\left[-\frac{\im_i}{m_0}\left(1-\frac{\sigma_i^2}{2m_0\im_i}\right)\right]\times \nonumber \\
&& \left\{1+{\rm erf}\left[ \frac{\im_i}{\sigma_i\sqrt{2}}
\left(1-\frac{\sigma_i^2}{m_0\im_i}\right)\right]
 \right\}\,.
\end{eqnarray}
If we want (as is the case for our data set) to implement data cuts that restrict the values of
$\im_i$ to be larger than some limiting value $m_l$, the likelihood of a single observation of a
modulation index $\im_i$ will be the expression above multiplied by a Heaviside step function, and
renormalized so that the likelihood $\ell_{i,\rm cuts}$ to obtain any value of $\im_i$ {\em above}
$m_l$ is 1:
\begin{equation}
\ell_{i,\rm cuts}[m_l] = \frac{H(\im_i-m_l)\ell_i}{\int _{\im_i=m_l}^\infty d\im_i \ell_i}\,.
\end{equation}
This renormalization enforces that there is no probability density for observed events ``leaking''
in the parameter space of rejected $\im_i$ values. In this way, it``informs'' the likelihood that
the reason why no objects of $\im_i<m_l$ are observed is not because such objects are not found in
nature, but rather because we have excluded them ``by hand.''

The integral in the denominator is analytically calculable,
\begin{eqnarray}
\int _{\im_i=m_l}^\infty d\im_i \ell_i &=&
\frac{1}{2}\left\{
\exp\left(\frac{\sigma_i^2}{2m_0^2}-\frac{m_l}{m_0}\right)\times \right.\nonumber \\
&& \left[1+{\rm erf}\left(\frac{m_l}{\sigma_i\sqrt{2}}
 -\frac{\sigma_i}{m_0\sqrt{2}}\right)\right]
 \nonumber \\
&& \left.+1-{\rm erf}\left(\frac{m_l}{\sigma_i\sqrt{2}}\right)
\right\}
\,.
\end{eqnarray}

The likelihood of $N$ observations of this type is
\begin{equation}\label{likelihood0}
\mathcal{L} (m_0) = \prod_{i=1}^N \ell_{i,\rm cuts}[m_l]\,.
\end{equation}

If we wish to study two parts of the $S_0$ parameter space with different cuts (as in, for example,
Figure~\ref{all_m_vs_s}, where we have a cut of $m_l = 0.02$ for $S_0>0.4~\mathrm{Jy}$, and a different cut of
$m_u=0.06$ for $0.06~\mathrm{Jy}\le S_0\le 0.4~\mathrm{Jy}$), we can implement this in a straight-forward way, by
considering each segment of the $S_0$ parameter space as a distinct ``experiment'', with its own
data cut. If the first ``experiment'' involves $N_l$ objects surviving the $m_l$ cut, and the second
``experiment'' involves $N_u$ objects surviving the $m_u$ cut, then the overall likelihood will
simply be
\begin{equation}\label{likelihood}
\mathcal{L} (m_0)= \prod_{i=1}^{N_l} \ell_{i,\rm cuts}[m_l]\prod_{i=1}^{N_u} \ell_{i,\rm cuts}[m_u]\,.
\end{equation}
Maximizing equation~(\ref{likelihood}) we obtain the maximum-likelihood value of $m_0$, $m_{0,\rm
  maxL}$. Statistical uncertainties on this value can also be obtained in a straight-forward way, as
equation~(\ref{likelihood}), assuming a flat prior on $m_0$, gives the probability density of the
mean intrinsic modulation index $m_0$ of the subset under study.

\subsubsection{Variability Analysis---Population Studies---Results}\label{popresults}

Here we apply the formalism introduced in \S~\ref{popstudies} to examine whether the intrinsic
modulation index $\im$ correlates with the physical properties of the sources in our sample. We will
be testing whether the distributions of $\im$-values in subsets of our monitoring sample split
according to some source property are consistent with each other. To verify that our analysis does
not yield spurious results, we first discuss two test cases where the likelihood analysis {\em
  should not } find a difference in the variability properties of the different subsets considered.

\begin{figure}[t]
  \includegraphics[width=\columnwidth, clip=]{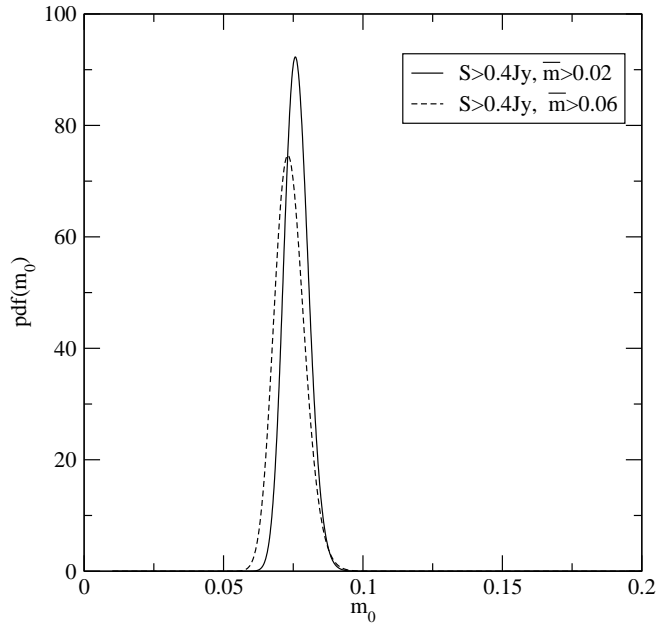}
  \caption[CutsNoCuts]{\label{cuts} Probability density of $m_0$ for the subset of bright CGRaBS blazars not
    found in 1LAC, for two values of the cutoff for data acceptance: $m_l =0.02$ (solid line), and
    $m_l = 0.06$ (dashed line). The two distributions are consistent with a single value. }
\end{figure}

\begin{figure}[t]
  \includegraphics[width=\columnwidth, clip=]{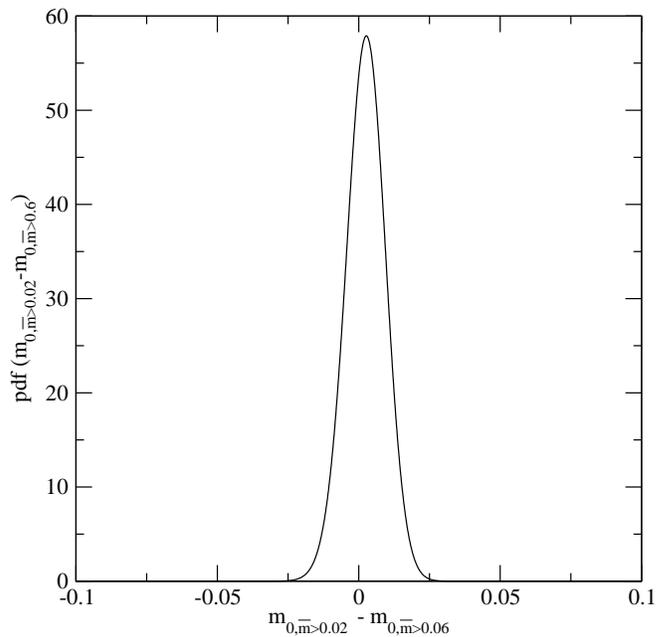}
  \caption[cc_cuts]{\label{cc_cuts} Probability density of the {\em difference} between the mean modulation
    index $m_0$ for the two sets considered in Figure~\ref{cuts}. The difference is consistent with
    zero within $1\sigma$. }
\end{figure}

The first case tests whether the data cuts discussed in \S~\ref{vanalysis} are implemented correctly
in \S~\ref{popstudies}. To this end, we calculate $\mathcal{L}(m_0)$ for the set of non
gamma-ray-loud CGRaBS blazars (blazars not found in 1LAC) in our monitoring sample with $S>0.4$ Jy,
in two different ways: first, by applying an $\im$ cut at $m_l = 0.02$; second, by applying an $\im$
cut at $m_l = 0.06$ (a much more aggressive cut than necessary for the particular bright blazar
population). The increased value of $m_l$ in the second case should not affect the result other than
by reducing the number of data points and thus resulting in a less constraining likelihood for
$m_0$. This is indeed the case, as we see in Figure~\ref{cuts}, where we plot the probability density
of $m_0$ for the two subsets. That the two distributions are consistent with each other is
explicitly demonstrated in Figure~\ref{cc_cuts}, where we plot the probability density of the {\em
  difference} between the means $m_0$ of the two subsets (which is formally equal to the
cross-correlation of their individual distributions). The difference is consistent with zero within
$1\sigma$.

\begin{figure}[t]
  \includegraphics[width=\columnwidth, clip=]{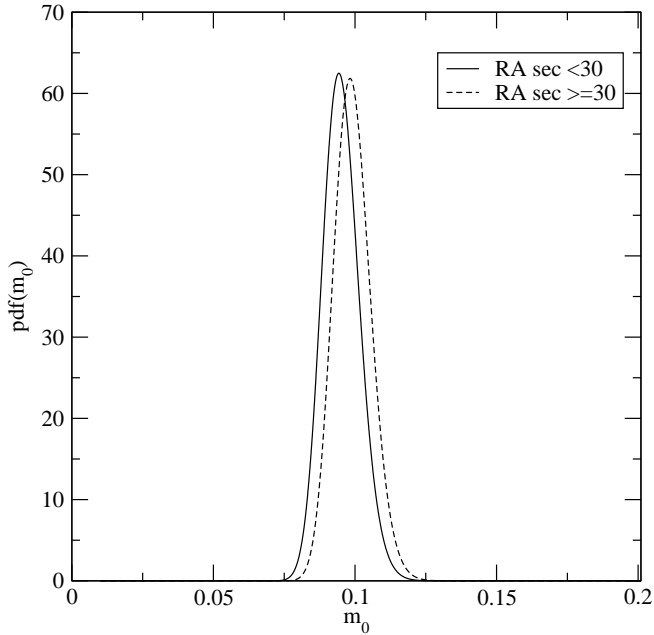}
  \caption[SmallBigSecs]{\label{secs} Probability density of $m_0$ for the subset of bright CGRaBS blazars: those
    with seconds of RA $<$~30~s (solid line) or $\ge$~30~s (dashed line).  The two distributions are
    consistent with a single value. }
\end{figure}

\begin{figure}[t]
  \includegraphics[width=\columnwidth, clip=]{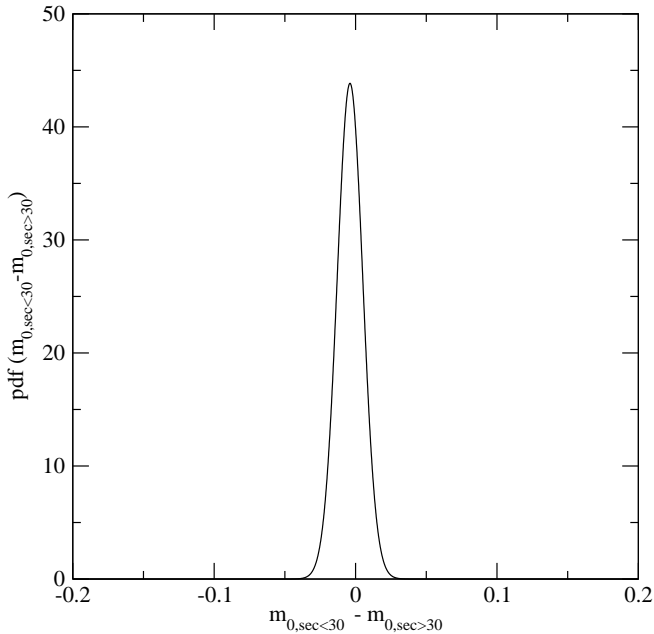}
  \caption[cc_secs]{\label{cc_secs} Probability density of the {\em difference} between the mean modulation
    index $m_0$ for the two sets considered in Figure~\ref{secs}. The difference is consistent with
    zero within $1\sigma$. }
\end{figure}

The second case tests whether a split according to a source property {\em without} physical meaning
and with the {\em same} value for the cutoff modulation index $m_l$ will yield probability densities
for the $m_0$ that are consistent with each other. For this reason, we split the population of
bright ($S>0.4$ Jy) CGRaBS blazars in our monitoring sample in two subsets in the following way: we
divide the RA of each source by 1~min.  If the remainder of this operation is $<30$~s, we include
this source in the first subsample (depicted by a solid line in Figure~\ref{secs}). If the remainder
is $\ge 30$~s we include the source in the second subsample (depicted by a dashed line in
Figure~\ref{secs}). As expected, the probability distributions of $m_0$ for the two subsamples,
shown in Figure \ref{secs}, are consistent with each other. This is also explicitly demonstrated in
Figure~\ref{cc_secs}, which shows the probability density of the difference between the $m_0$ in the
two subsamples. The difference is consistent with zero within $1\sigma$.

\begin{figure}[t]
\includegraphics[width=\columnwidth, clip=]{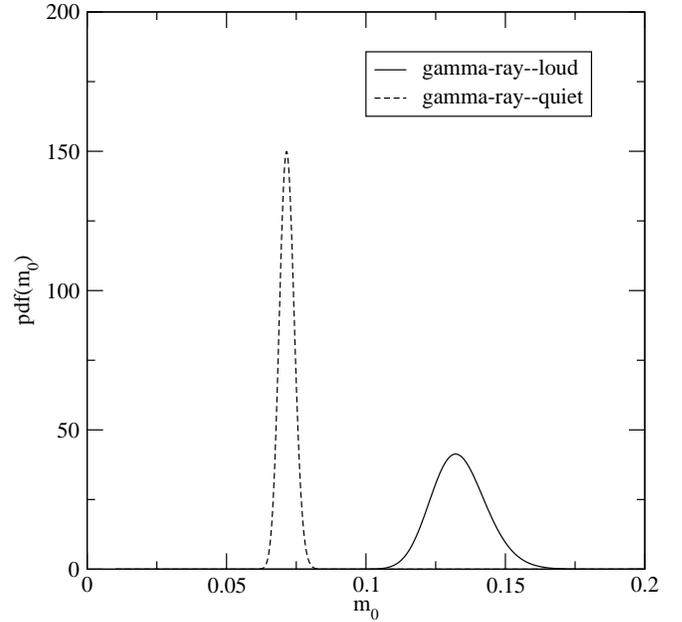}
\caption[NewGammaNonGamma]{\label{gamma} Probability density of $m_0$ for CGRaBS blazars in our monitoring sample that
  are (solid line) and are not (dashed line) included in 1LAC. The two distributions are {\em not}
  consistent with a single value. }
\end{figure}

\begin{figure}[t]
  \includegraphics[width=\columnwidth, clip=]{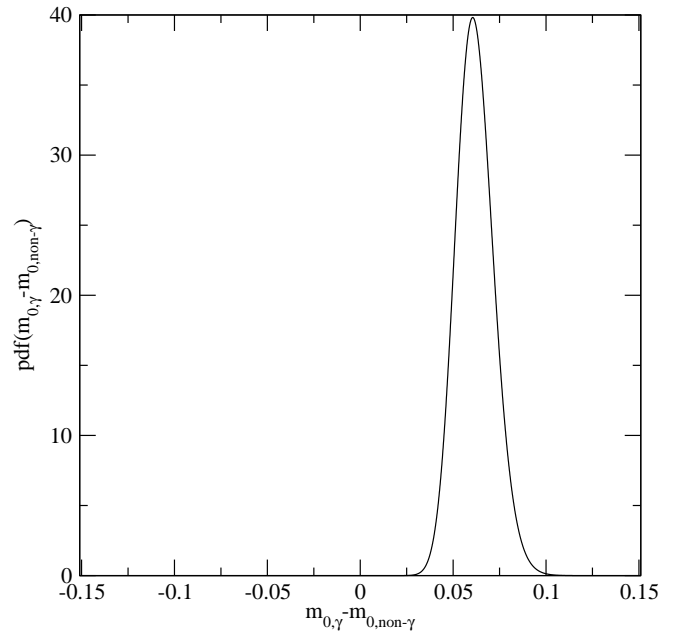}
  \caption[cc_gng]{\label{cc_gamma} Probability density of the {\em difference} between the mean modulation
    index $m_0$ for the two sets considered in Figure~\ref{gamma}. The peak of the distribution is
    $7\sigma$ away from zero. }
\end{figure}

\begin{figure}[t]
  \includegraphics[width=\columnwidth, clip=]{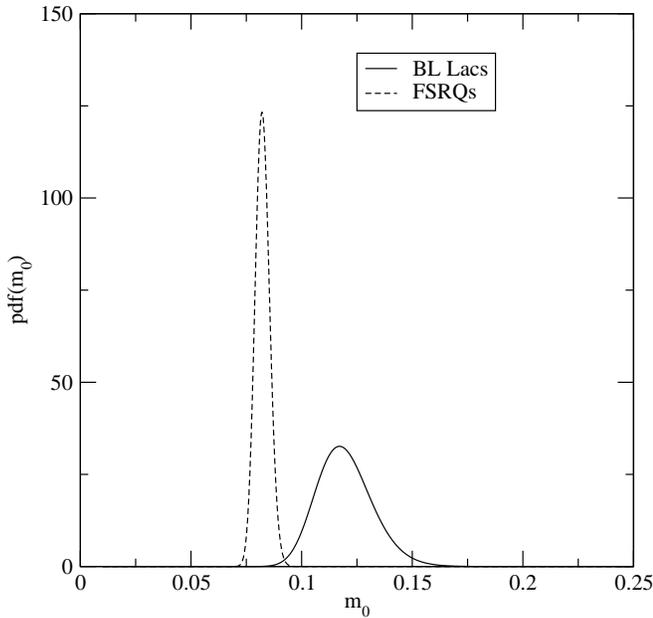}
  \caption[BLL_FSRQ]{\label{bf} Probability density of $m_0$ for BL~Lac (solid line) and FSRQ (dashed line)
    CGRaBS blazars in our monitoring sample. The two distributions are {\em not} consistent with a
    single value. }
\end{figure}

\begin{figure}[t]
  \includegraphics[width=\columnwidth, clip=]{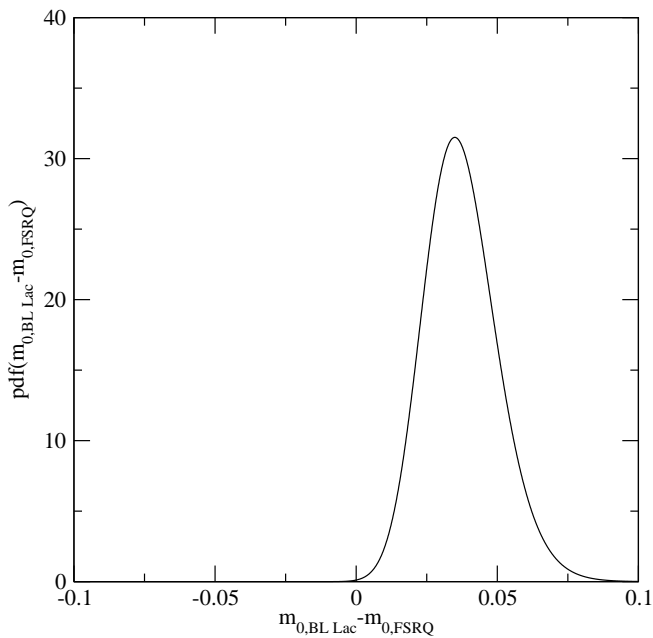}
  \caption[cc_bf]{\label{cc_bf} Probability density of the {\em difference} between the mean modulation
    index $m_0$ for the two sets considered in Figure~\ref{bf}. The peak of the distribution (at
    $0.035$) is more than $3\sigma$ away from zero. }
\end{figure}
We next examine subsets defined according to physical properties of the sources. The first criterion
we apply is whether the source has been detected by \emph{Fermi}-LAT at a significance level high
enough to warrant inclusion in the 1LAC catalog. For sources with $S_0<0.4~\mathrm{Jy}$ we apply a
cut $\im>m_u=0.06$ and for sources with $S_0\ge 0.4~\mathrm{Jy}$ a cut $\im>m_l=0.02$. The results
are shown in Figures~\ref{gamma} and~\ref{cc_gamma}. The set of sources that are included in 1LAC is
depicted by a solid line, while the set of sources that are not in 1LAC is depicted by a dashed
line. The two are not consistent with each other at a confidence level of $7\sigma$
(Figure~\ref{cc_gamma}), with a maximum-likelihood difference of 6 percentage points, with
gamma-ray-loud blazars exhibiting, on average, a higher variability amplitude {\em by almost a
  factor of 2} versus non gamma-ray-loud blazars.

We also examine the variability amplitude properties as a function of optical spectral
classification. We analyze the subsets of CGRaBS BL~Lacs and FSRQs. The probability densities for
the mean $m_0$ of the two subsets are shown in Fig \ref{bf}. The results for BL~Lacs (FSRQs) are
plotted as a solid (dashed) line. The two curves are not consistent with each other---the BL~Lacs
appear to have, on average, higher variability amplitude than the FSRQs. We verify this finding by
plotting, in Figure~\ref{cc_bf}, the probability density of the difference between the $m_0$ of
BL~Lacs and FSRQs. The most likely difference is $3.5$ percentage points, and it is more than
$3\sigma$ away from zero. Note that the difference between BL~Lacs and FSRQs is less significant
than that between gamma-ray-loud and non gamma-ray-loud blazars. This is both because the most
likely difference in $m_0$ values between the BL~Lac and FSRQ subsets is smaller and because the
BL~Lac sample is smaller than the gamma-ray-loud blazar sample: only 94 BL~Lacs satisfy the data
cuts we impose, versus 190 gamma-ray-loud blazars.  As a result, the constraints on the intrinsic
distribution of modulation indices (i.e. on $m_0$) are stronger in the latter case.
\begin{figure}[t]
  \includegraphics[width=\columnwidth, clip=]{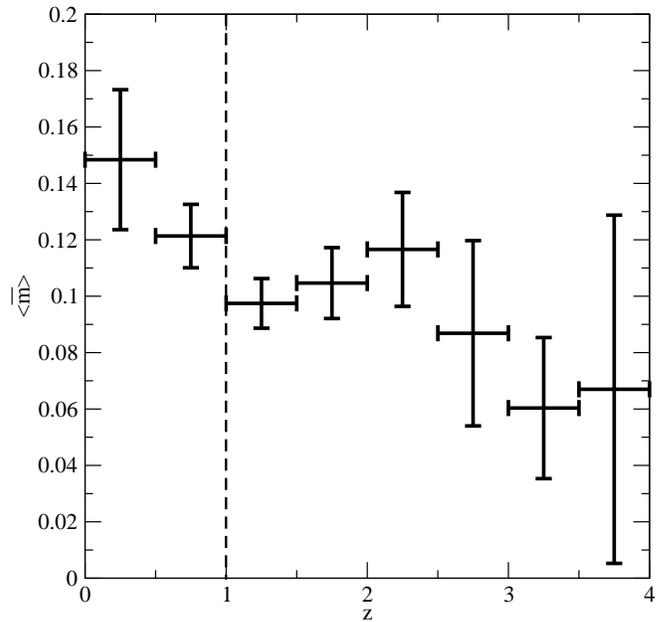}
  \caption[m_vs_z]{\label{mvsz} Mean $\im$ in redshift bins of $0.5$ for bright ($S>0.5~\mathrm{Jy}$) FSRQs
    in our monitoring sample.}
\end{figure}

\begin{figure}[t]
  \includegraphics[width=\columnwidth, clip=]{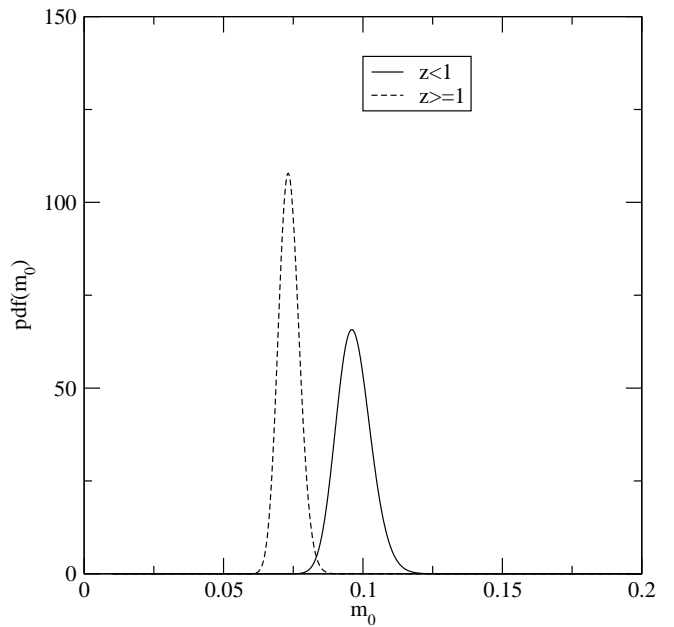}
  \caption[Lowz_Highz]{\label{thez}Probability density of $m_0$ for FSRQs in our monitoring sample with $z<1.0$
    (solid line) and $z\ge1.0$ (dashed line) . The two distributions are {\em not} consistent with a
    single value.}
\end{figure}
\begin{figure}[t]
  \includegraphics[width=\columnwidth, clip=]{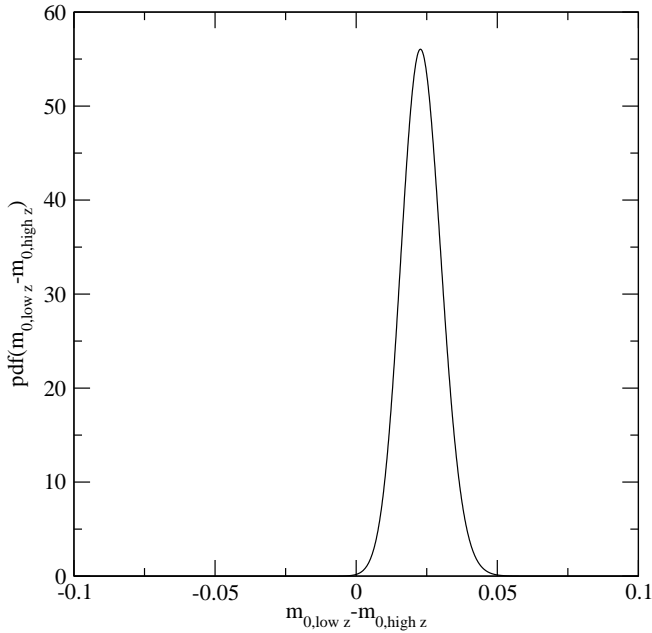}
  \caption[cc_z]{\label{cc_thez} Probability density of the {\em difference} between the mean modulation
    index $m_0$ for the two sets considered in Figure~\ref{thez}. The peak of the distribution (at
    $0.023$) is more than $3\sigma$ away from zero. }
\end{figure}

Finally, we examine the dependence of variability amplitude on redshift. In Figure~\ref{mvsz} we
plot the mean $\im$ (as calculated by a simple average rather than the likelihood analysis) in
redshift bins of $\Delta z = 0.5$ for bright ($S\ge0.4~\mathrm{Jy}$) FSRQs with known redshifts in
our monitoring sample. We exclude BL~Lacs from this analysis so as not to bias the result, as
BL~Lacs with known redshifts are located at low $z$, and we have also already shown that they have a
higher mean $\im$ compared to FSRQs. Although the errors are large, there is a hint of a trend
toward decreasing variability amplitude with increasing redshift. We further test the significance
of this result by splitting sources in our monitored sample in high- and low-redshift subsets with
the dividing redshift at $z=1$ (dashed line in Figure~\ref{mvsz}). In the two subsets we also
include faint ($S<0.4~\mathrm{Jy}$) sources, with the usual cut at $m_u=0.06$. The probability
density for the mean $m_0$ of each subset is shown in Figure~\ref{thez}, where the solid curve
corresponds to low-redshift blazars and the dashed curve to high-redshift FSRQs. We find that
low-redshift FSRQs have {\em higher}, on average, intrinsic modulation indices. The result is shown
to be statistically significant in Figure~\ref{cc_thez}, where we plot the probability density of
the difference between $m_0$ in each subset. The most likely difference is found to be about 2.5
percentage points, and more than $3\sigma$ away from zero.

\section{Discussion and Conclusions}

We have discussed in detail the OVRO 40-m telescope 15~GHz monitoring program. We have presented
results from the first two years of observations, including reduced data and light curves for all
sources in our monitoring sample.

We have derived the variability amplitude properties of all blazars in our sample through a
likelihood analysis that deconvolves the intrinsic variability from scatter induced due to errors in
individual flux density measurements and accounts for uncertainties due to finite (and different) sampling
in each source to calculate an {\em intrinsic modulation index} as well as uncertainties on its
value. We have used these intrinsic modulation indices to study whether and how the variability
amplitude is correlated with physical properties of our sources.

We have found that the distribution of intrinsic modulation indices is different between sources
that have / have not been detected by Fermi in GeV gamma rays; between BL~Lacs / FSRQs; and between
FSRQs at high and low redshifts.

Our most significant result is that gamma-ray-loud sources have a higher, on average, variability
amplitude, as quantified by the intrinsic modulation indices, than non-gamma-ray-loud sources. The
most likely difference in mean modulation index is about 6 percentage points, so that
gamma-ray-loud sources have, on average, a variability amplitude {\em almost a factor of 2 higher}
than sources not found in 1LAC. The result is very significant statistically, with the
maximum-likelihood difference being $7\sigma$ away from 0.

It is not clear whether a selection effect or an intrinsic difference is responsible for this
deviation between the two subsets. It is, for example, conceivable, that {\em all} CGRaBS blazars
are potentially gamma-ray-loud at some part of their activity cycle and, given enough observation
time, all of them would enter their ``flaring'' state (that would presumably be characterized by
enhanced broad-band luminosity, including increased flux density at 15~GHz) and would be detected in
GeV gamma rays. If this is the case, then the blazars that have been detected by Fermi so far would
be the ones that happened to have been in their ``flaring'' state during the first year of Fermi
operations, and it would be expected that they are seen to have a higher, on average, variability
amplitude in 15~GHz as well. In this scenario, given more time, more blazars in our sample will
enter at some point their ``flaring'' state; they will be detected in gamma rays, and the amplitude
of their 15~GHz emission will also increase. If we were to repeat the same experiment after another
two years of observation, the source numbers in the two subsamples would change, but not the average
population properties: more sources would be detected in gamma rays, but these sources would now
also exhibit a higher $\im$. The average $m_0$ of each population would not change appreciably, but
sources would move from one category (non-gamma-ray-loud) to the other (gamma-ray-loud).  If on
the other hand we have seen all blazars in our monitoring sample in all activity states, then the
variability amplitudes of each are not expected to change appreciably if we observe them for longer,
and the number statistics in the two categories will likely remain fixed (for a fixed gamma-ray flux
detection threshold). In this scenario, the variability amplitude is the result of some intrinsic,
persistent physical property of blazars, which is also related to the gamma-ray activity of the
source.

BL~Lacs are found to have higher, on average, intrinsic modulation indices than FSRQs, by about
$3.5$ percentage points. Due to the smaller-number statistics and on-average smaller difference
variability, the difference is less significant, but still more than 3$\sigma$ away from 0. In
addition, among our FSRQ subsample, low redshift sources are found to have higher, on average,
intrinsic modulation indices, with the most likely difference on the mean being about 2.5 percentage
points, also more than $3\sigma$ away from 0.

The latter difference is not easy to interpret as an indication of source evolution, as there are
competing effects that could affect the result in either way. On the one hand, sources at higher
redshift have been observed for a shorter rest-frame time interval due to time-dilation effects, so
it is conceivable that high-redshift sources have not been followed through their complete activity
cycles, and their intrinsic modulation indices will increase as they are observed for longer. On the
other hand, sources at higher redshift are being observed at a higher rest-frame frequency. Because
the radio variability amplitude increases with increasing frequency, this effect should yield higher
modulation indices for higher-redshift sources. As our monitoring program is continued, the
importance of the first effect will decrease (or, conversely, with long enough light-curves, we
could select to look at shorter light curve segments for our low-redshift sources, corresponding to
the same rest-frame time interval as for our highest-redshift sources). The second effect is not
affected by length of observation time, however it operates in the opposite direction to the
observed effect.  Should the trend persists as it is seen here (low-redshift sources have higher
modulation indices), this might be an indication of source evolution with cosmic time toward higher
variability amplitudes.

We have found larger variability for the LAT-detected blazars, for the BL~Lac-type blazars, and for
the sources at lower redshift. In addition, we have noted the relatively modest overlap between the
(EGRET-like) CGRaBS sources and the early LAT detections. These facts may be related: it has already
been shown in~\citet{abdo_first_2010} that the decrease in LAT effective area below 0.3~GeV has
strongly biased the LAT detections to the relatively hard-spectrum high-peak blazars, especially the
BL~Lacs, compared to the EGRET sample. Indeed 1LAC contained $\sim$~50\% BL~Lacs, while for EGRET
FSRQs outnumbered BL~Lacs by $>3\times$. These BL~Lacs are radio fainter and tend to be lower-power
sources at relatively low redshift. Thus we expect from the higher variability amplitude found for
BL~Lacs and lower-$z$ sources in this paper that the LAT-detected blazar sample should have higher
average variability. We note however that the difference in variability amplitude between
gamma-ray-loud and gamma-ray-quiet blazars in our sample is much larger than the difference between
BL~Lacs and FSRQs, so this effect cannot be attributed in its entirety to different BL~Lac/FSRQ
number ratios in the CGRaBS and LAT-detected blazar samples.  As LAT exposure increases and as
refinement of the event cuts allows more effective area at lower energy, we might expect an increase
in high-power, high-redshift FSRQ detections, with steeper gamma-ray spectral indices. These sources
would have variability of lower amplitude and/or longer observed timescale. Indeed, continued LAT
exposure is detecting more CGRaBS sources and we expect that our OVRO-monitored sample will allow an
excellent comparison of radio variability statistics on rest-frame timescales comparable to those
now probed for the nearby BL~Lacs.

In conclusion, we have, for the first time, been able to explicitly demonstrate that the radio
variability amplitude of blazars exhibits positive correlations with physically meaningful
properties of the sources. Our findings are important steps toward understanding the physical
differences between blazars with otherwise similar properties which, however, differ in their
gamma-ray activity.

The variability amplitude in radio frequencies has never before been considered as a differentiating
property between blazar samples; this was largely due to practical purposes, as never before has
such a large, preselected, statistically complete sample been monitored for as long a time and with
as high a cadence. The compilation of the CGRaBS sample \citep{healey_cgrabs:all-sky_2008}, for
example, was based on radio flux density, radio spectral index, and X-ray flux;
variability information was not included, not because it was not considered important, but rather
because such information was, at the time, unavailable. As a result, the CGRaBS catalog had only
moderate success in predicting sources that would emerge as gamma-ray sources in the LAT
era. However, by providing a preselected sample defined by robust statistical criteria, the CGRaBS
sample has allowed us to make unprecedented progress in studying the population properties of
blazars, as in this work.

As the additional, non-CGRaBS blazars that have been discovered by Fermi have now been added to our
monitored source sample, our program will allow us to confirm and expand these results in upcoming
years. In addition, by establishing, through the results of this work as well as those presented in
\cite{fluxfluxmethodology} and \cite{fluxfluxLAT}, that there is a close connection between
gamma-ray and 15 GHz blazar emission, we are justified to expect that additional progress in blazar
jet physics is to be expected through cross-correlations in the time domain between 15 GHz and
\emph{Fermi}-LAT gamma-ray light curves. Such cross-correlations will be discussed in an upcoming publication
\cite{max-moerbeck_ovro_2010}.

\acknowledgements We are grateful to Russ Keeney for his tireless efforts in support of observations
at the Owens Valley Radio observatory.  The OVRO 40-m monitoring program is supported in part by
NASA grants NNX08AW31G and NNG06GG1G and NSF grant AST-0808050.  Support from the Max-Planck
Institut f\"{u}r Radioastronomie for upgrading the OVRO 40-m telescope receiver is also
acknowledged. OGK acknowledges the support of a Keck Institute for Space Studies Fellowship. WM
acknowledges support from the US Department of State and the Comisi\'{o}n Nacional de
Investigaci\'{o}n Cient\'{i}fica y Tecnol\'{o}gica (CONICYT) in Chile for a Fulbright-CONICYT
scholarship.  VP acknowledges support for this work provided by NASA through Einstein Postdoctoral
Fellowship grant number PF8-90060 awarded by the Chandra X-ray Center, which is operated by the
Smithsonian Astrophysical Observatory for NASA under contract NAS8-03060, and thanks the Department
of Physics at the University of Crete for their hospitality during the completion of part of this
work.

{\it Facilities:} \facility{OVRO:40m}

\bibliography{ms}

\end{document}